# *Modular Pulse Synthesizer for Transcranial Magnetic Stimulation with Flexible User-Defined Pulse Shaping and Rapidly Changing Pulses in Sequences*

*Z. Li, J. Zhang, A. V. Peterchev, and S. M. Goetz*

## Abstract

The temporal shape of a pulse in transcranial magnetic stimulation (TMS) influences which neuron populations are activated preferentially as well as the strength and even direction of neuromodulation effects. Furthermore, various pulse shapes differ in their efficiency, coil heating, sensory perception, and clicking sound. However, the available TMS pulse shape repertoire is still very limited to a few biphasic, monophasic, and polyphasic pulses with sinusoidal or near-rectangular shapes. Monophasic pulses, though found to be more selective and stronger in neuromodulation, are generated inefficiently and therefore only available in simple low-frequency repetitive protocols. Despite a strong interest to exploit the temporal effects of TMS pulse shapes and pulse sequences, waveform control is relatively inflexible and only possible parametrically within certain limits. Previously proposed approaches for flexible pulse shape control, such as through power electronic inverters, have significant limitations: Existing semiconductor switches can fail under the immense electrical stress associated with free pulse shaping, and most conventional power inverter topologies are incapable of generating smooth electric fields or existing pulse shapes.

Leveraging intensive preliminary work on modular power electronics, we present a modular pulse synthesizer (MPS) technology that can, for the first time, flexibly generate high-power TMS pulses (~ 4,000 V, ~ 8,000 A) with user-defined electric field shape as well as rapid sequences of pulses with high output quality. The circuit topology breaks the problem of simultaneous high power and switching speed into smaller, manageable portions, distributed across several identical modules. In consequence, MPS TMS can use semiconductor devices with voltage and current ratings lower than the overall pulse voltage and distribute the overall switching of several hundred kilohertz among multiple transistors. MPS TMS can synthesize practically any pulse shape, including conventional ones, with fine quantization of the induced electric field. Moreover, the technology allows optional symmetric differential coil driving so that the average electric potential of the coil, in contrast to conventional TMS devices, stays constant to prevent capacitive artifacts in sensitive recording amplifiers, such as electroencephalography (EEG).

MPS TMS can enable the optimization of stimulation paradigms for more sophisticated probing of brain function as well as stronger and more selective neuromodulation, further expanding the parameter space available to users.

## Introduction

Magnetic stimulation is a technique to noninvasively stimulate neurons in the brain of conscious humans, and also in the spine or the peripheral nervous system [1-7]. Transcranial magnetic stimulation (TMS) causes neurons to fire action potentials in response to strong brief current pulses that generate magnetic fields reaching across the skull and inducing electric fields in a relatively focal target location [8, 9]. In addition to such *writing* of artificial signals into the brain, certain rhythms and patterns can further modulate neuronal circuits, i.e., influence how circuits process endogenous signals they receive [10]. TMS has become a key tool in experimental brain research and is widely used in medical diagnosis and treatment [11-14]. It is, for example, FDA-cleared for the treatment of depression, obsessive compulsive disorder, smoking addiction, and migraine as well as for cortical mapping [15-24]. TMS is also under investigation for many other disorders [1, 11, 25-27].

The spatial focality is largely determined by the TMS coil design and can be on the order of 0.5–1 cm³, corresponding to approximately a million neurons [28-30]. Although TMS is typically called focal, it simultaneously activates neurons of different types and functions across an entire gyral crown—including inhibitory and excitatory neuron populations as well as projections from other circuits [31-37]. Thus, a pulse can even simultaneously activate neurons that counteract each other. Presumably as a consequence of this limited selectivity, the neuromodulatory and therapeutic effects of conventional TMS protocols are relatively weak and variable [15-18, 38-41]. In experimental brain research, the weak and variable effects necessitate large subject populations to establish sound group effects. In treatment, however, it limits the efficacy on the individual level.

Latest coil designs bring the spatial selectivity, i.e., focality, to its limits [9]. There is mounting evidence that the temporal shape of TMS pulses affects the functional selectivity of neural stimulation as well as the direction and strength of neuromodulation [37, 42-59]. For example, monophasic TMS pulses produce stronger and more selective neuromodulation than biphasic pulses [60-65]. However, the way conventional technology generates monophasic pulses is highly inefficient. The second half of the current pulse is generated through resistive damping so that the entire pulse energy is converted into heat (see Figure 1) [66]. In consequence, existing TMS devices cannot generate conventional monophasic pulse trains at rapid rates, with few exceptions that were never deployed widely due to their excessive power demand [8, 67]. The use of standard monophasic pulses, as just one example of alternative waveforms, is limited to few applications and pulse protocols. Clinical treatments, however, are exclusively performed with so-called biphasic pulses, which have inferior efficacy and selectivity [18, 68-71].

Other pulse shapes have been shown to provide activation selectivity, increase the neuromodulation efficacy of pulse trains with that pulse shape, or affect physical features of stimulation, such as coil heating or sound emission. Various asymmetric near-rectangular pulses were shown to increase selectivity and produce stronger neuromodulation effects with similarities to conventional monophasic pulses [42]. Varying the pulse symmetry and pulse duration furthermore apparently allows shifting of the activation balance between different neuron populations [43-47, 56, 58, 59, 72, 73].

Stimulation with pulses of different durations demonstrated that pulse shapes also affect a subject's or patient's perception of the pulse on the scalp, likely due to the different activation dynamics of nociceptors and other sensory fibers in the skin compared to the various cortical neurons [74]. Pulses with the majority of their spectral content in higher frequency ranges emit less sound, which is more than just a technical nuisance and artifact of TMS as it concur-rently stimulates auditory circuits [75-80]. The loud clicking sound of pulses could previously not be isolated from the electromagnetic stimulation and is always exactly in sync with it. This confounds TMS studies and clinical applications because auditory stimulation is known to have a strong neuromodulatory effect on brain stem and/or neocortical circuits [76, 81].

Beyond the physiology, optimized pulse shapes allow reducing the heating of TMS coils, which traditionally limits repetitive TMS protocols and has to be kept within certain limits for safety reasons [82-85].

Finally, the ability to change the TMS pulse shape can enable analysis of the stimulated neuron population and its activation characteristics, e.g., through identifying the first order of its dynamics, which, beyond distinguishing between various targets, may allow diagnosis of various diseases [72, 86, 87].

Regrettably, the temporal aspect of the TMS stimulus cannot be explored and optimized fully because existing technology has fundamental limitations of the ability to control the shape and sequences of the induced electric field pulses. To exploit potentially increased selectivity, neuromodulation strength, and energy efficiency of optimized stimulus waveforms, there is a need for a TMS device that allows flexible synthesis of pulse shapes and sequences. An unconstrained experimental search and application of optimal waveforms with respect to activation selectivity, strength, and reliability of various neuromodulation techniques, as well as physical aspects, such as pulse sound emission or coil heating, would require a device and circuit technology that allows the generation of practically any pulse shape and any sequence.

## *Existing Device Technology*

### *Available devices and pulse shapes*

In conventional TMS devices, the coil voltage pulse, and hence the electric field waveform, has a damped cosine shape [88]. Most commercial devices offer a selection of biphasic and monophasic sinusoidal current pulses with very limited control over the pulse shape and width [49, 89-91]. These devices exclusively implement high-voltage oscillator circuits, which are charged up slowly and release the energy rapidly through oscillation between a pulse capacitor and the stimulation coil, activated by a high-voltage high-power switch (Figure 1A, B) [88, 90, 92]. The oscillation is either terminated typically after one period to retrieve some of the pulse energy (biphasic TMS) or over-damped to zero, in which case the pulse energy is dissipated as heat (monophasic TMS). In contrast to monophasic TMS, biphasic devices do not intentionally use any damping resistor to shape the pulse and can recycle a large share of the energy in the pulse capacitor after a pulse and only replenish the loss through a power supply [93]. Changing the pulse shape requires rewiring of the circuit, e.g., by increasing the capacitance or adding damping resistors, but is technologically cumbersome and practically yields only a handful of pulse options with limited repetition rates [49, 52, 57, 94-97].

We previously developed controllable pulse parameter TMS (cTMS) to address some of these limitations of conventional TMS devices [95, 98-100]. cTMS and related approaches, such as flexTMS [101, 102], use switching between one or two oscillators with large capacitors to enable electric field pulses with nearly rectangular shape and some control over the pulse width and, in some cases, the amplitude ratio between the positive and negative pulse phases (Figure 1C) [95, 99, 103]. Coupling various capacitors and inductors has been suggested to bring pulse-shape flexibility [104-107]. However, these devices are still very limited in their pulse shapes and also cannot generate conventional sinusoidal pulses, such as the traditional monophasic or biphasic waveforms, for comparative studies or to perform already approved clinical procedures [95, 98, 99, 101].

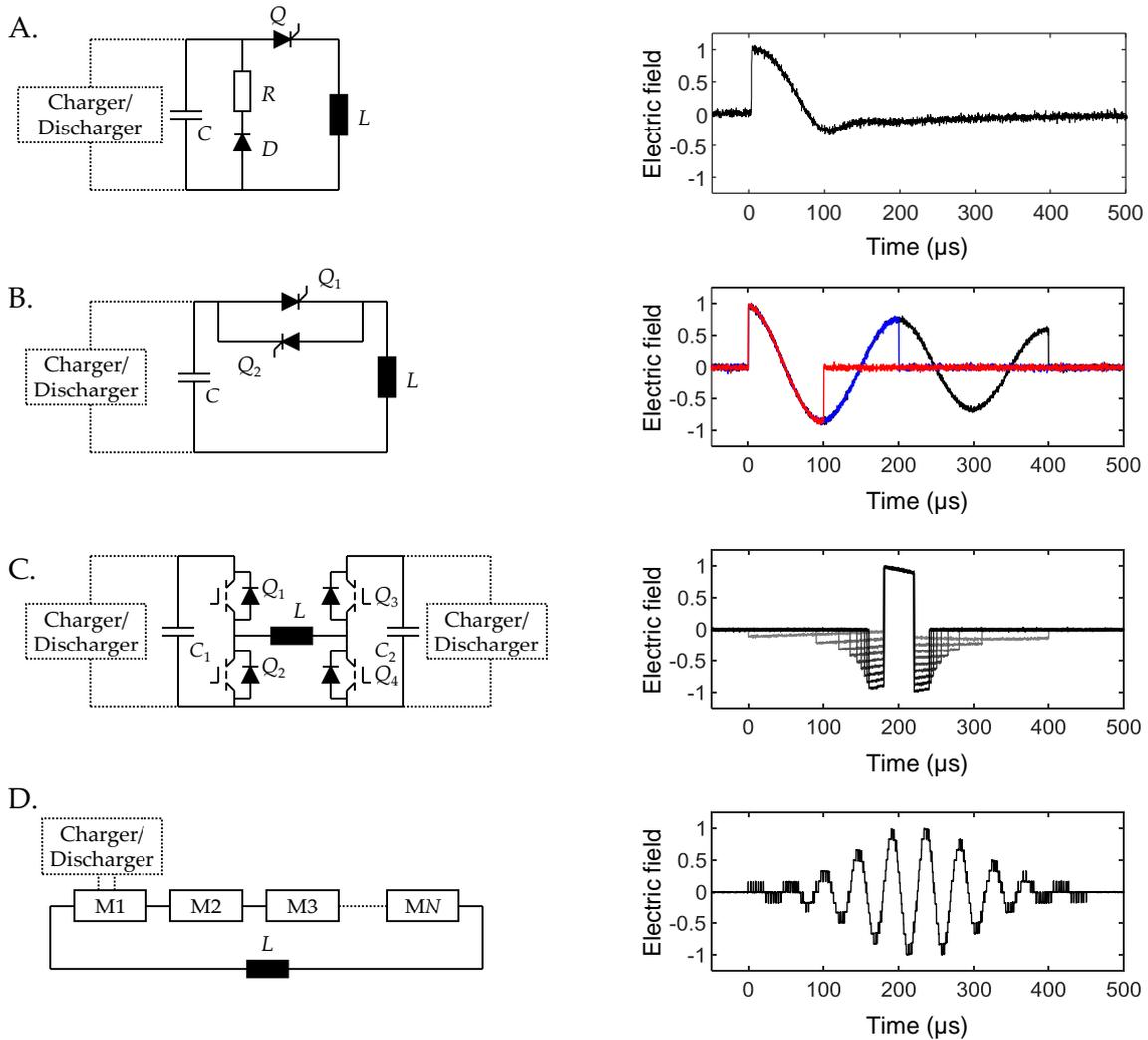

**Figure 1.** Circuit topologies (left column) and corresponding typical pulse shapes (right column) of various TMS technologies: (a) Monophasic stimulator (typically with thyristors), (b) bi- and polyphasic stimulator (existing with thyristors and IGBTs), (c) example of a bridge-based TMS device using IGBTs, representing also all other bridge approaches, and (d) modular pulse synthesizer.

These technological limitations of existing TMS devices stem from several factors. Fundamentally, high energy and power levels as well as fast pulse dynamics are required for transcranially evoking action potentials in cortical neurons through electromagnetic induction. This necessitates TMS devices to employ high voltages (typically up to 3,000 V), large currents (up to 8,000 A), and fast switching (< 1 µs) [2, 5, 52, 92, 108-110]. To handle these extreme requirements, existing TMS devices exploit simple but inflexible circuit topologies, usually one or several oscillators, and high-power but slow semiconductor switches (thyristors in conventional TMS and insulated-gate bipolar transistors (IGBT) in later devices) [99, 111-114].

In the devices discussed above, the pulse shape or family of pulse shapes are still very characteristic for the specific stimulator circuit. A practical technology to generate relatively freely almost any existing and user-defined pulse shape would have to use a single flexible and efficient circuit topology. For example, if that technology could generate pulses without the need for deliberate damping with resistors, it would allow the generation of monophasic pulses with energy loss and coil heating comparable to biphasic pulses [82].

*Limitation in combining various pulses in sequences flexibly and efficiently*

No available single TMS device allows controlled changes of the pulse amplitude and pulse shape between two or more pulses delivered in rapid succession. Such rapid changes are, for example, relevant for paired-pulse measurement protocols, conditioning pulses, or interleaved modulation and probing pulses [50, 115-118]. Several devices can be combined to feed a single coil with pulses to rapidly change the pulse shape or just the amplitude from one pulse to another. However, for every different pulse shape class or amplitude involved, a separate device is necessary. The inability to alter the pulse amplitude rapidly results from the circuit topologies of conventional devices, which do not allow rapid changes of the voltage of the energy storage capacitor that determines the pulse amplitude. Such limitations also apply to existing TMS devices using various bridge topologies (e.g., Figure 1C), including a restricted set of pulse shapes that can be sustained in repetitive trains as many pulses transfer charge from one capacitor to the other; inability to combine different pulse shapes or amplitudes in a sequence or a train; and suboptimal energy losses and coil heating [95, 98, 101]. In conclusion, there are multiple technological limitations of existing devices that impede the exploration or adoption of stimulation protocols that could enhance TMS applications.

*Limitations for generating any user-defined pulse*

Various approaches have been implemented or proposed to increase the flexibility of pulse shaping by relatively rapid switching of the coil during the pulse. Some used conventional voltage-source inverter circuits with bridge configurations of transistors, typically IGBTs [119], which are known for example from motor inverters of electric vehicles [120, 121]. The bridge circuit switches the voltage of the coil rapidly between a small number of voltages (typically positive supply, negative supply, and zero voltage) to generate an approximation of a reference current shape. Current-source alternatives were considered but would provide less flexibility in TMS [122, 123]. Half-bridge or so-called chopper arrangements with switching modulation circuit-wise have large similarities to the first version of cTMS [99, 100]. In contrast to full bridge configurations, they switch between two voltages (on and off) only to generate pulses with merely one current polarity.

The use of power electronic inverters was accordingly suggested and tested early in TMS as an apparently obvious solution for the generation of practically any current shape. The concept of power electronic inverters for TMS has been revisited multiple times since then [124, 125]. However, the concept has serious problems. Few transistors have to manage three challenges at the same time: the full current, the maximum voltage, and the full bandwidth, i.e., frequency content and dynamics of the pulse, which overwhelm practically any available power semiconductor. In short, the IGBTs of a conventional inverter have to switch the full pulse currents in the kiloampere range fast and frequently while they need to block the entire voltage. In conventional TMS devices, the semiconductor switches—typically thyristors—are overloaded with currents multiple times their continuous ratings for the short duration of a pulse. Overloading is not only a means to save cost, but it reduces the parasitic properties of the semiconductor devices, such as the capacitance and other charge-storing effects as well as associated reverse recovery, and therefore allows faster switching speed. IGBTs can also be overloaded, but not as far as thyristors. Furthermore, the stress of TMS-level currents on transistors is enormous when they have to perform hard commutation at the same time. High-voltage IGBTs are moreover rather slow compared to the spectral content of TMS pulses and lossy in switching. Consequently, in contrast to motor inverters, where the switching frequency is orders of magnitude higher than the generated waveforms, IGBTs can only switch a few times per TMS pulse.

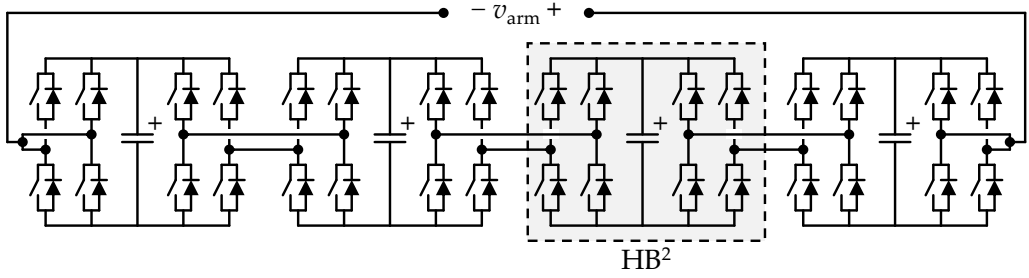

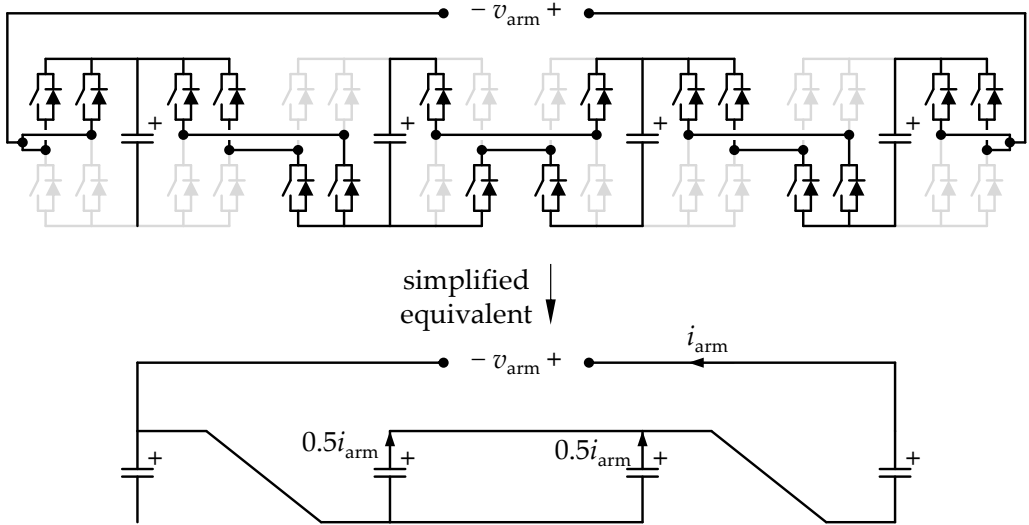

**Figure 2.** A. Circuit topology of the power train with here four four-bridge modules. B. Example for a state of the system with two modules including their internal capacitor connected in series and two in parallel to generate two output steps. Two-bridge modules operate similarly, but the parallel mode would have to be replaced by bypass. See Figure 3 for definition of all states.

Most importantly, the current shape, which such inverters control as required for motors and which appears to look smooth, is not directly relevant for the stimulation effect of TMS. Instead, the stimulation effect of TMS is mediated by the induced electric field. The electric field is proportional to the coil current rate of change and hence approximately follows the coil voltage time course. Since the TMS coil voltage switching is performed between the available supply voltage levels, which usually are in the kilovolt range, the electric field is in stark contrast to the apparently smooth current a high-amplitude rectangular pulse waveform. Consequently, the neuronal membrane potential, which is driven by the induced electric field, experiences a high-rate burst of ultra-brief sharp-edged pulses with varying pulse duration in the range of microseconds or tens of microseconds, each pulse in the burst with maximum stimulator amplitude. Since the active neuronal membrane is nonlinear, such pulses have a different physiological impact than pulses with a more continuous electric field waveform, such as those of conventional TMS pulses. Indeed, the strong nonlinearity of neurons is relevant particularly for short pulse durations and high amplitudes [126-129]. Thus, while novel TMS pulses with high-frequency content may be an interesting parameter space to explore, such inverter circuits cannot reproduce standard TMS pulses or synthesize other pulse waveforms with a continuous electric field profile. We previously argued that a conventional inverter circuit cannot provide at the same time the quality, power, and bandwidth necessary for flexible TMS pulse synthesis, and introduced alternative circuits; these circuits split the high voltage, current, and switching frequency into smaller, manageable portions distributed over many faster, lower-power semiconductor devices [130-136].

In this paper, we present a modular pulse synthesizer TMS device (MPS TMS) that uses a modular circuit topology combined with the latest fast-switching unipolar field-effect transistors using wide-bandgap silicon-carbide semiconductors. The latter exhibit less slow charge-storage effects than silicon IGBTs or super-junction field-effect transistors, permitting rapid switching at high power. The combination of a modular circuit topology and fast semiconductor switches enables a TMS device that can generate practically any pulse shape, including the existing ones, with high output quality of the induced electric field. The high pulse shape flexibility allows generating user-defined pulses with only a few constraints, such as a maximum voltage, current, and duration, as well as the current returning to zero at the end of the pulse. Furthermore, since the synthesis of pulses becomes a digital control problem, and therefore a software task, practically any set of various pulse shapes, durations, directions, and amplitudes can be combined into rapid pulse sequences without the need to mechanically reconfigure the circuit or slowly change the charge level on capacitors between pulses as in available devices. For example, MPS TMS can synthesize conventional monophasic pulses accurately with energy recovery after a pulse to enable high efficiency and also high repetition rates—a feature that has been technologically limited to biphasic pulses [130]. Furthermore, since MPS TMS can generate existing pulses with high fidelity, it could equivalently generate FDA-approved treatment protocols.

## *Modular Circuit Topology*

TMS pulses challenge power electronics due to the simultaneous occurrence of high power (currents of typically 5–10 kA, high voltage up to ~ 3 kV) and high dynamic range (≫10 kHz with low distortion at least > 100 kHz; often desired sharp transients in the pulse requiring notably more), which requires fast switching or circuit reconfiguration for free control. In conventional circuits, such as choppers, half bridges, or full bridges, individual power semiconductors have to manage these three at once. Although available semiconductor technology—such as high-frequency high-electron-mobility transistors for fast switching in the MHz range but insufficient voltage and current capability, or high-voltage high-current IGBTs for MVA of power but large parasitic inductances and capacitances as well as slow bipolar charge-carrier dynamics—could achieve each of the three requirements alone, to date no semiconductor type can satisfactorily manage all of those at the same time. However, even if new devices, such as wide-bandgap semiconductor IGBTs, may allow it in the future, the rectangular electric field would, as outlined above, generate mostly rectangular pulse bursts. Without a large high-power filter between the power electronics and the coil, these bursts would not resemble the desired smooth pulse or be equivalent to any conventional pulse shape.

We combined two ingredients to solve this problem. First, we designed a circuit class that splits power and switching into smaller, manageable portions [130-133]. Figure 1D depicts the circuit topology as a string of modules, which can even be kept identical. The smaller units contain their own capacitors and transistors, which only need to deal with a fraction of the pulse voltage, and can be bridge circuits. Second, we adopted fully unipolar field-effect silicon-carbide (SiC) transistors. These SiC transistors avoid the slow dynamics of minority carriers in pn junctions of bipolar devices such as thyristors and IGBTs as well as the large charge-storing effect in the voluminous drift zone of high-current super-junction silicon field-effect transistors and their large diode reverse recovery due to necessarily high doping levels [137-141]. Slow IGBTs or superjunction transistors in the modules could limit switching or circuit reconfiguration to only few times during a TMS pulse.

Accordingly, a number of lower-voltage modules contribute their individual lower power and lower switching dynamics to generate a more powerful and faster output for a TMS pulse. The lower voltage in each module allows the use of lower-voltage components for capacitors and transistors with faster switching dynamics. At the same time, if appropriately coordinated, the overall switching is likewise distributed and temporally shifted among the individual modules so that each module only needs to deal with a fraction of the switching speed to achieve overall effective switching rates approaching the MHz range. Similar to the voltage, the individual modules further only change the output electric field in small steps so that they have fine control over the field amplitude in stair-case fashion, which switching modulation can smoothen further. Due to the low amplitude and high frequency of the steps, the energy content of the distortion product is small. Compact filters can eliminate such distortion if needed.

Each individual module forms a local dc voltage supply with a storage capacitance and contains several transistor bridges. The transistor bridges connect the module to the neighboring ones or the stimulation coil (see Figure 2). The capacitors are charged before a pulse starts. The transistors allow rapid reconfiguration of the overall circuit: by appropriate activation and deactivation of transistors, the module capacitors can be included into the coil-driving circuit with either polarity or bypassed, effectively increasing or reducing the output voltage supplied to the stimulation coil within a microsecond. A circuit with $N$ modules and similar capacitor voltages can step the output voltage with $2N + 1$ steps, i.e., $-N, \ldots, 0, \ldots, N$. As in any TMS device, the current is almost entirely inductive and therefore follows approximately the time integral of the output voltage. Any switching modulation to generate intermediate steps can be performed exclusively with the size of one voltage step. The small steps are a fundamental advantage of the proposed configuration in contrast to the afore-mentioned conventional power inverters.

The module circuit preferably contains four individually controllable transistor bridges in each module so that a module is connected to each of the adjacent modules through two interconnections. This pairwise interconnection allows module capacitors not only to be dynamically connected in series to the other included capacitors but also in parallel for balancing charge or sharing load [132]. The extra bridges appear to double the number of transistors needed. However, all transistors share load and the circuit achieves the same semiconductor utilization. Thus, the individual transistors in the four transistor bridges, each of which nevertheless aggregates several parallel transistor dies, can have only half the current rating so that the overall power semiconductor amount is the same for two or four transistor bridges, while the opportunities for control increase [142]. Furthermore, the bridges can act in parallel so that the four-bridge module can operate equivalently to a two-bridge module with the exact same performance. Table I compares these options of the technology.

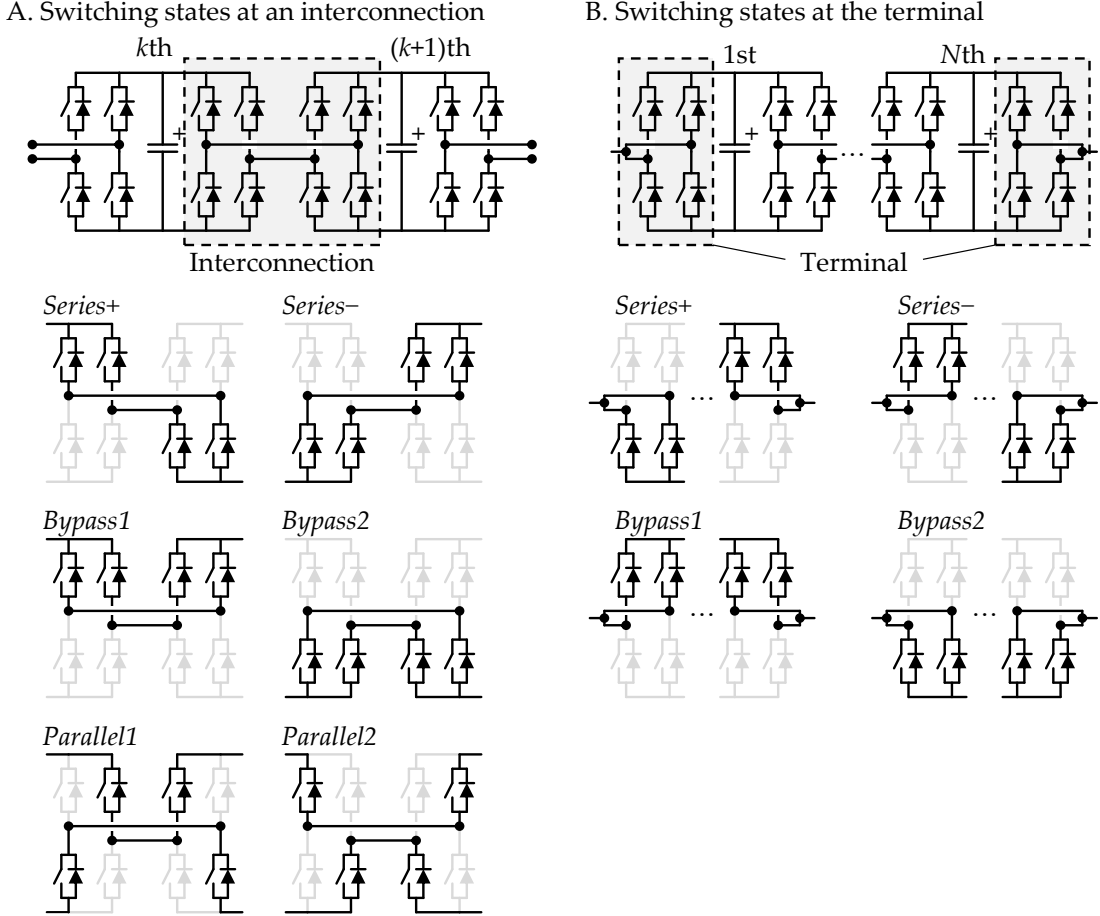

**Figure 3.** Switch states of an interconnection A between two modules or B the outmost modules connecting to the coil. The states influence how the capacitors of the modules are electrically connected relative to their neighbors. Two-bridge modules can generate only series and bypass states, four-bridge modules additionally parallel circuit configurations between modules. Series states increase the voltage in positive or negative direction. Parallel modes distribute the current load and balance charge across the capacitors. With fast transistors, the transition from one state to another can occur on the order of 100 ns and several hundred times per millisecond.

The overall circuit is furthermore symmetric. Although the device can generate pulses the same way as conventional stimulators, the circuit symmetry further allows generating pulses differentially. Differential pulse generation means that the machine feeds the two coil terminals with inverse voltages (relative to ground) so that the average electric potential of the coil stays close to zero throughout the pulse. Conventional devices, in contrast, generate pulses exclusively in a single-ended style, where one coil terminal is fixed, often grounded, and the pulse voltage that drives the current through the coil is solely generated through the other terminal. As a consequence, the average electrical potential of the coil fluctuates throughout a pulse with rather sharp edges. Such electrical potential fluctuations on the order of thousands of volts within microseconds turn the coil into an antenna and generate strong capacitive interference in sensitive electronics nearby. This interference is part of the stimulus and therefore synchronous with it. Thus, it leads to artifacts in recording equipment for neural signals, particularly electroencephalography (EEG), which is not easily separable and often saturates the amplifier so that the amplifier is insensitive to neural signals right after a stimulus [143].

The modular circuit allows various ways for recharging both on the module level with charging each module independently or only a few of them in combination with a charge distribution through the power loop. Alternatively, charging can also occur entirely on the system level by distributing a voltage to the module string through the power loop. The implementation described below charges through a dc supply to the capacitor of one module, from which the power is supplied to the others via the parallel channel offered by the four-bridge modules.

Table I. Comparison between module alternatives

| | Two Bridges | Four Bridges |
|---|---|---|
| Topology | 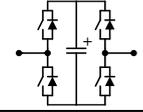 | 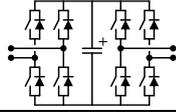 |
| Modulation range | $-1 \leq m \leq 1$ (four-quadrant operation) | |
| Sensorless-balancing | No | Yes |
| Discrete switch count | 4 | 8 |
| Switch current rating | $i_{arm}$ | $\tfrac{1}{2} i_{arm}$ |
| ON-state resistance[†] | $\tfrac{1}{2} r_{on}$ | $r_{on}$ |
| Switch conduction loss | $i^2_{arm} r_{on} + 2 V_F i_{arm}$ | |
| Capacitor conduction loss | $i^2_{arm} r_{cap} m$ | $\tfrac{1}{2} i^2_{arm} r_{cap} m \sim i^2_{arm} r_{cap} m$ |

## *Control and Balancing*

The modules collectively and dynamically reconfigure the series and parallel configuration of module capacitors presented to the stimulation coil. The various options of how modules can interact and form the dynamically changing overall circuit is best represented and controlled by so-called module or interconnection states [144]. The four-bridge module topology has four interconnection states, specifically series plus (i.e., stepping up the voltage in positive direction) and series minus (i.e., stepping down the voltage in negative direction), which insert the module's capacitor into the coil loop in series with the others with either polarity, bypass, which disconnects the module capacitor from the pulse current by guiding the current around, passive, where all transistors are turned off so that only the free-wheeling diodes act as a rectifier from the module terminals towards the module capacitor, and various parallel states, of which anyone is sufficient to connect the capacitor in parallel to one or more other module capacitors (Figure 3). Two-bridge modules can only generate the subset of series plus, series minus, bypass, and passive, while the parallel configuration is not possible.

In contrast to conventional TMS devices, the control has to implement several more complex functions. It has to translate a specified continuous electric field or current shape into a quantized representation that uses the fine amplitude granularity and temporal switching speed of the system (modulator), decide which transistor in the system has to switch on or off at which point in time (scheduler), and balance the load so that each module is delivering similar charge, similar power, and a similar share of the switching load [145].

In the first approximation, the output voltage $v_{coil}(t)$ is the sum of the module capacitor voltages $v_{m,k}$ that are in the series plus state minus those in series minus per

$$v_{coil}(t) = \sum_k s_k v_{m,k}(t), \qquad (1)$$

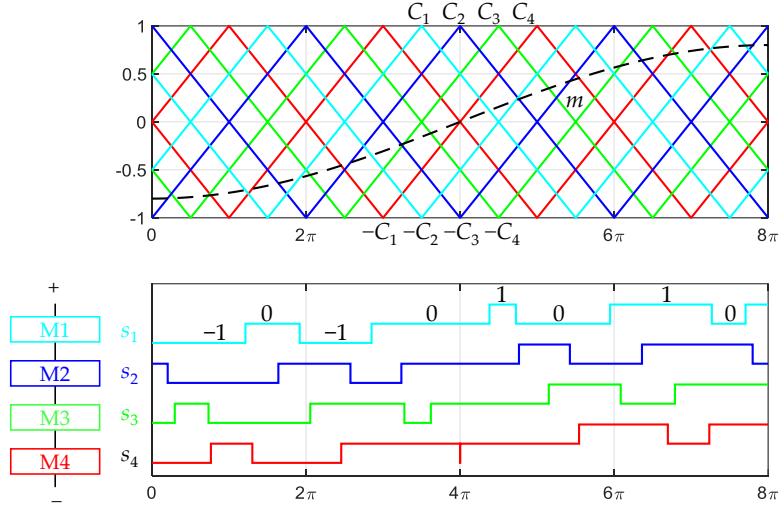

**Figure 4.** Principle of phase-shifted carrier modulation (PSC) with four modules $M_k$ and accordingly four carriers $C_k$. to reproduce the dashed black reference curve $m(t)$.

assuming $s_k(t)$ represents the state for the $k$-th module or interconnection according to

$$s_k = \begin{cases} +1 & \text{(Series +) if } m \geq C_k, \\ -1 & \text{(Series −) if } m \leq -C_k, \\ 0 & \text{(Parallel or Bypass) otherwise.} \end{cases} \quad (2)$$

The modulation method we preferably use is based on phase-shifted carrier modulation, PSC (Figure 4), which can perform modulation, scheduling, and balancing with appropriate design [146, 147]. The switching signals are determined by comparing a pulse shape reference $m(t)$ against a set of $k$ cyclic triangular carriers $C_k(t)$ shifted in phase relative to each other to interleave switching of the modules for a high effective switching rate. Each corresponds to a module. The comparison result (+1, −1, or 0) is conveniently the switching state $s_k$. The PSC modulation determines the switching rate by its carrier frequency. Since the carriers are evenly shifted in phase, the effective switching rate at the TMS output is $N f_\text{carrier}$, e.g., with $f_\text{carrier}$ = 50 – 100 kHz. PSC rotates states through all modules so that the modules get loaded similarly. Proper choice of the carrier sequence can furthermore exploit the opportunities of the parallel mode for charge balancing across modules as well as to reduce conduction loss [146, 148].

As a first approximation, the pulse shape reference $m$ can be set in an open-loop control approach as $m(t) = v_\text{coil}(t)/(N v_\text{m})$ for a desired coil voltage profile $v_\text{coil}(t)$ and $N$ modules with equal and constant module capacitor voltage $v_\text{m} = v_{\text{m},k}$. If the reference is given as a coil current $i_\text{coil}^*$, setting

$$m = \frac{1}{N v_\text{m}} L \frac{d i_\text{coil}^*}{dt} \quad (3)$$

can produce a very close result to the reference (see Figure 5A). However, during the pulse, the capacitor voltages can change and deviate from each other, while the inner resistance $R_\text{i}$ of the circuit, cables, and the coil absorbs some of the output voltage before it reaches the inductance of the stimulation coil and therefore distort the pulse.

Conventional closed-loop control during a pulse, which would be the solution in conventional power electronics to compensate such deviations, is not recommended as extremely fast feedback would be necessary for the current and voltage dynamics of TMS. Instead the pulsed application allows a closed-loop control scheme from pulse to pulse to adjust parameters. If detailed circuit parameters are known, e.g., measured beforehand, we can pre-compensate the change of the module dc capacitors and the stray resistance in $m$, as in

$$m = \frac{1}{N v_\mathrm{m}(t)} \left( i^*_\mathrm{coil} R_\mathrm{i} + L \frac{d i^*_\mathrm{coil}}{dt} \right). \tag{4}$$

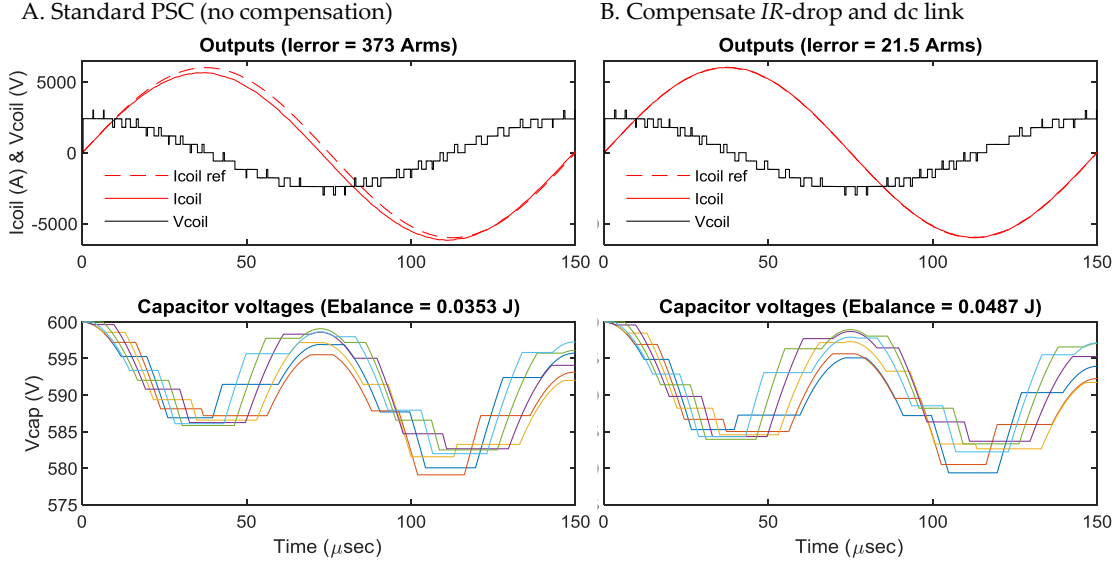

**Figure 5.** Simulation results (A) without compensation and (B) with a reference compensating resistive voltage drop and module capacitor voltage variations. The model does not use the parallel mode but only series and bypass to demonstrate how the capacitors drift apart. Parallelization clears such voltage differences.

Equation (4) adds the ohmic voltage drop to the modulation reference and considers the discharging effects of the module capacitor voltages, which are represented by the time dependence of $v_\mathrm{m}(t)$. For simplicity, $v_\mathrm{m}(t)$ models the trajectory of the average module capacitor voltage throughout the pulse, which is predicted from energy perspective (capacitor energy $e_C$, coil energy $c_\mathrm{coil}$, Joule loss $e_R$, and output energy $e_\mathrm{out}$) assuming an ideal output current as in

$$\begin{cases} v_C = \sqrt{2 e_C} / \sqrt{NC}, \\ e_\mathrm{out} = e_\mathrm{coil} + e_R = \frac{1}{2} L i^{*\,2}_\mathrm{coil} + \int_0^t i^{*\,2}_\mathrm{coil} R\, d\tau, \\ e_C = e_C(0) - e_\mathrm{out}. \end{cases} \tag{5}$$

The improvement is obvious when compared to the uncompensated standard PSC as is shown in Figure 4 ($L_\mathrm{coil}$ = 9 μH, $R_\mathrm{i}$ = 36 mΩ, $C_m$ = 3.6 mF, $N$ = 6, no parallel state). Both PSC methods leave a noticeable module voltage spread on the order of 5 V at the end of the pulse as the control does not use the parallel mode there. Parallelization, as uniquely offered by the four-bridge modules, clears such voltage differences. It can generate loss ($E_\mathrm{loss} = ¼\, C_\mathrm{m}\, \Delta V_\mathrm{m}^2$) but if done regularly, e.g., instead of every bypass state during a pulse, can be negligible [132]: both examples produce < 50 mJ balancing loss while the overall energy loss can exceed 50 J. Model-predictive control, which estimates the future during control, can further reduce loss and deviation from the reference, but does not appear necessary for fundamental operation [149].

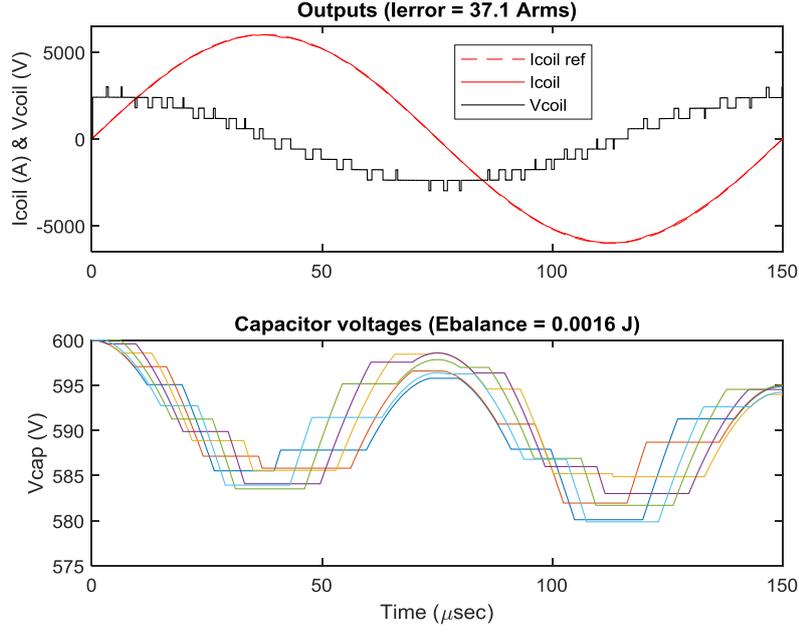

**Figure 6.** Simulation results using a reference that compensates the resistive voltage drop, module capacitor voltage variations, and the final module voltage spread.

We furthermore added a formalism that allows adding balancing to the control, which can eliminate also the small balancing loss. The individual references of the modules are modified with additive modifiers per

$$m_k = m + \Delta m_k \quad \left(\sum \Delta m_k = 0 \text{ and } \max|\Delta m_k| \ll 1\right)$$

to achieve even closer module voltages than the balancing capabilities of PSC modulation. The constraints of zero sum and low amplitude for the modifiers $\Delta m_k$ are set such that they cause minimum distortion on the output. The most convenient setting for $\Delta m_k$ gives them the shape of the load current, but scales them differently according to the difference of the expected value of the final dc link voltage $v_k$ from the mean value, as in

$$\begin{cases} \Delta m_k = \eta_k i_{\text{coil}}^{\text{å}}, \\ \eta_k = \Delta Q_k \Big/ \int_0^T i_{\text{coil}}^{\text{å}\,2} dt, \\ \Delta Q_k = C\left(v_k - \tfrac{1}{N}\sum_{j=1}^{N} v_j(T)\right). \end{cases}$$

Thus, the composite modulation reference ($m + \Delta m_k$) exactly subtracts the specific charge difference $\Delta Q_k$ from Module $k$. Figure 6 demonstrates the balancing improvement.

As the device can generate practically any pulse shape, including user-defined pulses, the control has to enforce constraints that are naturally fulfilled in conventional technology due to its limitations. In addition to maximum current amplitude and charge levels per phase, these constraints also include that a pulse does not end with non-zero current. As the current is the solution of a differential equation representing the circuit excited by the modular pulse synthesizer, the control has to estimate the current at the end of the pulse ahead of time to avoid substantial residue magnetic energy in the stimulation coil associated with nonzero current at the end of the pulse. The passive state of the modules or module interconnection can serve to recycle such residual energy at the end of a pulse [130]. The passive state uses the diodes of the modules so that residue current of any polarity is rectified into the modules to charge the capacitors. It can be actively controlled in its voltage amplitude by replacing series states with passive and combining them with bypass or parallel states statically or preferably in switching modulation to set the discharge time course.

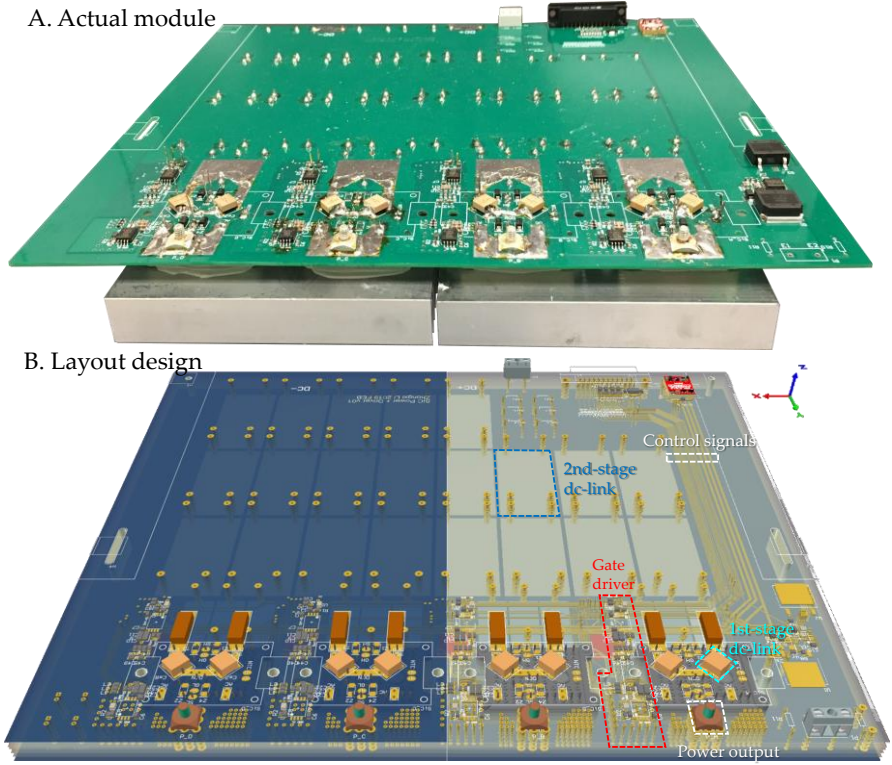

**Figure 7.** Circuit board layout (A. actual photo and B. design) with switching cells in the front and capacitance in the back. The transistors are on the bottom side, the driver and first-stage capacitance on the top.

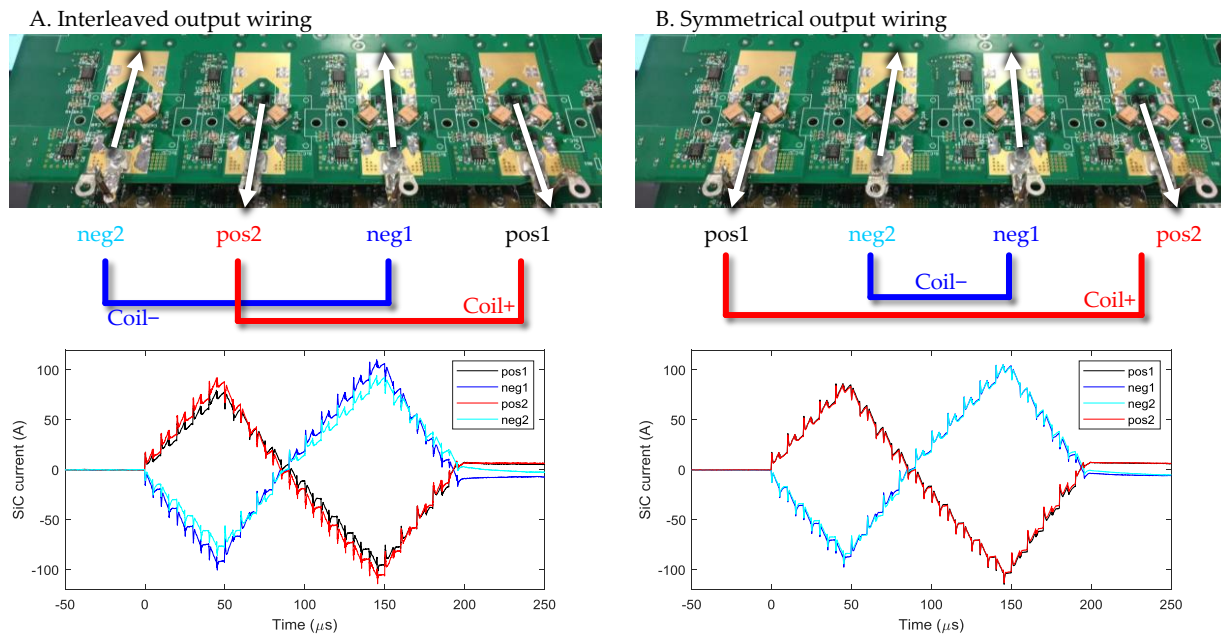

**Figure 8.** Two different assignments of four switching cells (discrete pair of high-side and low-side transistor) to two output bridges on one side of a module demonstrating the balance of the current among discrete parallel switching cells. Each switching cell contains discrete transistors (on the back, not visible), gate driver (black ICs between the golden islands), and first-level ceramic capacitance (yellow blocks in V configuration) to absorb the immediate energy demand and/or supply when the parasitic magnetic fields of the commutation current paths is discharged or charged.

## *Implementation*

We implemented the technology with six modules, each with four power half bridges to have access to parallel module interconnection states. Each module was implemented with 24 discrete SiC transistors (FF11MR12W1M1, Infineon, Germany, with press-fit circuit-board connection) to enable peak currents of at least approximately 8,000 A and voltages up to at least 4,000 V. The transistor and the included dies were tested for their channel saturation current level for various gate voltages to evaluate the transient over-load potential. In contrast to conventional TMS devices, the power circuit did not use a hand-wired setup of discrete components but heavy-copper printed circuit boards (PCB) that represent the identical modules. The four-layer PCBs ($2 \times 75$ µm and $2 \times 105$ µm copper) contain the SiC transistors, gate drivers, three levels of capacitors to allow fast switching without large voltage spikes, and module-level control electronics (see Figure 7). The PCBs are furthermore designed to fit into 19" racks. The pulse current is routed on the PCBs on all four copper layers in a laminated style for minimum parasitic inductance.

As the most critical component, we thoroughly optimized the switching cell, i.e., the high- and low-side transistors of a bridge together with their physical current routing. We placed a first-stage module capacitance and the gate driver in close vicinity with minimum inductance and coupling to enable rapid switching of the high TMS pulse currents. Following the switching cells, the entire module capacitor bank is split into three parts or stages with increasing distance and therefore parasitic inductance. The first stage capacitance, as part of the switching cells, is formed by low-inductance, low-resistance surface-mount ceramic capacitors (B58031U9254M062, EPCOS/TDK, see Figures 7 and 8) with 12 µF per module and provides and/or absorbs the immediate energy needed to discharge the magnetic field of the current path on the PCB before switching and to charge the magnetic field after switching. To further minimize those energies, the current path is routed for minimum commutation stray inductance, i.e., such that the current before switching and the current after switching share almost the same path and direction so that they can take over the magnetic field of each other. The individual switching cells were assigned to output bridges of the module in an interleaved manner for minimum inductance and best current balancing performance (see Figure 8). Gas discharge tubes (CG21000L, Littelfuse, Illinois, USA) across the module dc bus (and thus the capacitors) protect the switching cells against overvoltage and can absorb high energy levels.

The gate drivers (1EDI60H12AH, Infineon) are mounted on the other side of the PCB, underneath the corresponding transistors and supplied by compact dc/dc converters (MGJ1D121905MPC-R7, Murata, Japan). We switched the transistors with gate–source voltages of +19 V/–5 V to achieve (1) minimum conduction loss in the transistors, (2) fast turn-on/-off transients, (3) suppression of spurious turn-on events in response to capacitive as well as inductive interferences, and most importantly (4) to shift the transistor channel saturation to allow higher current levels. The gate and corresponding source signals were routed as parallel differential pair with minimum distance to the transistors with less than 25 mm trace length and connected with a Kelvin connection to the transistors. The Kelvin connection avoids large potential shifts in the gate-drive reference due any voltage drops generated by the large load current.

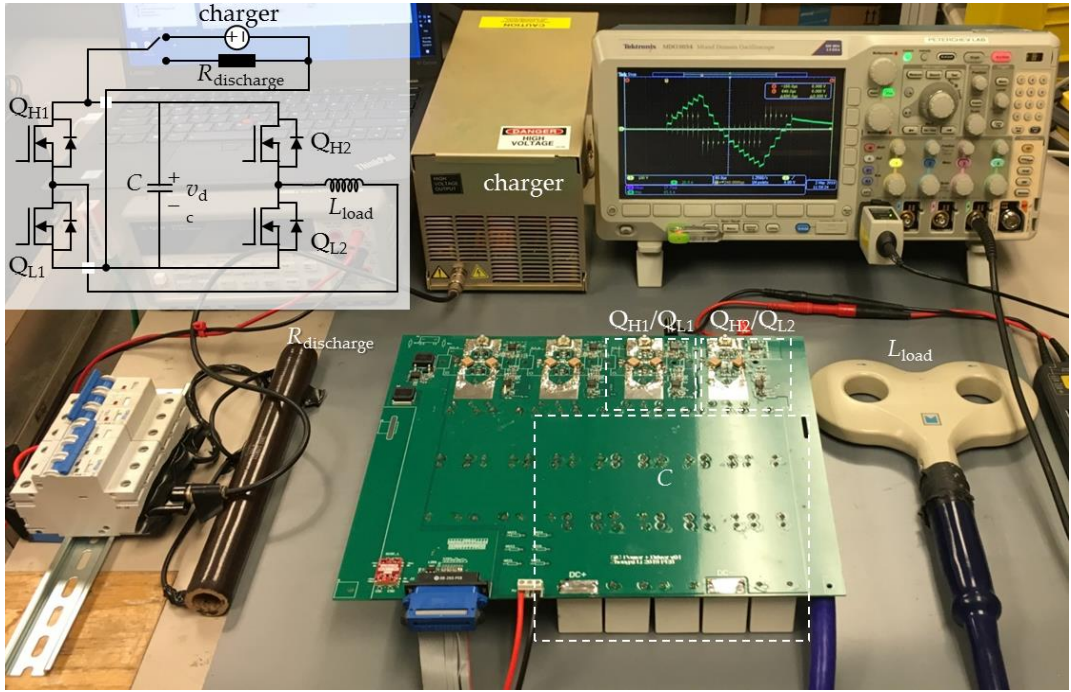

**Figure 9.** Double-pulse setup.

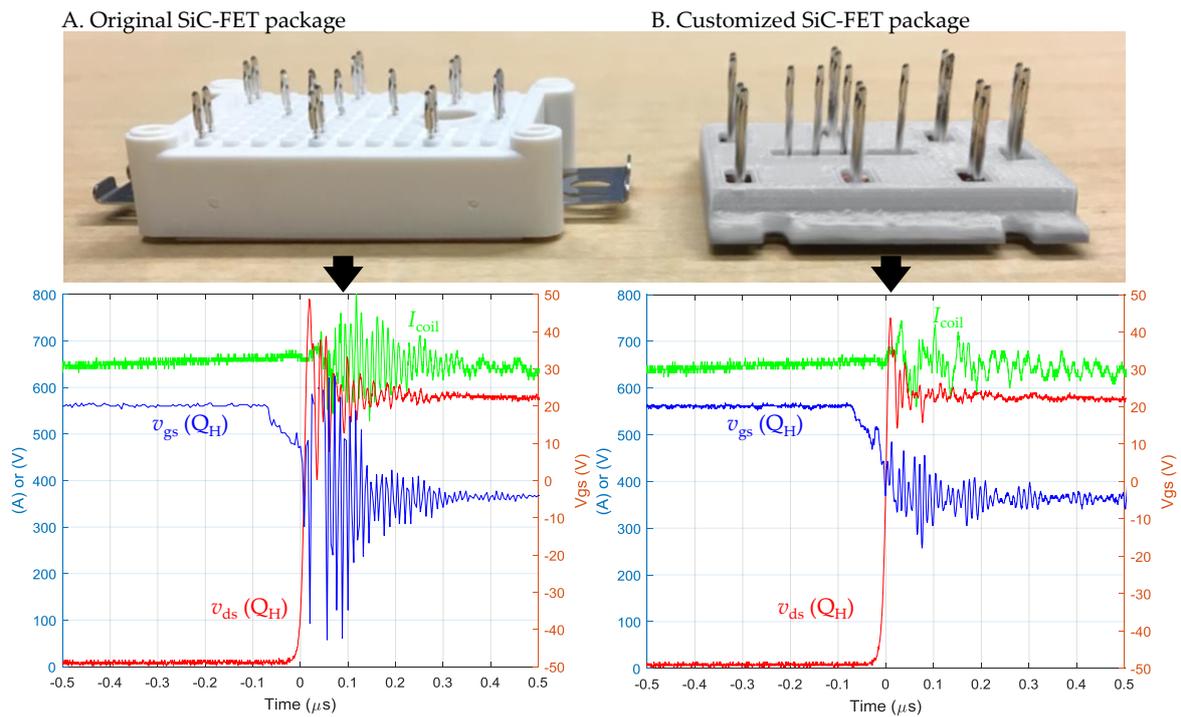

**Figure 10.** Example for one of four commutation conditions for A the original FF11 transistor bridge in the optimized layout and B with an optimized package for lower stray inductance. In the measurements, one transistor module is singled out.

The second-stage capacitance is formed by low-inductance metalized film capacitors with 194 µF per module (C4AQQBU4270A1XJ, Kemet, South Carolina, USA). The third-stage capacitance with 1.8 mF per module is farthest away from the switching cells and provides the major pulse energy (C4AQQBW5300A3MJ, Kemet). Each module further contains a discharge unit, which can decrease the module voltage and also rapidly bring the device into a safe state (i.e., below safety extra-low voltage, SELV) in case of unexpected events or failures. The discharge unit contains high-reliability normally-on reed relays (DBT71210U, Cynergy3 Components, England, UK) and discharge power resistors (AP101 470R J, Ohmite Manufacturing, Illinois, USA). The controller only opens the relays to allow charging after a successful self-test. The modules are connected with copper bars and alternate their position for close-to-ideal current sharing.

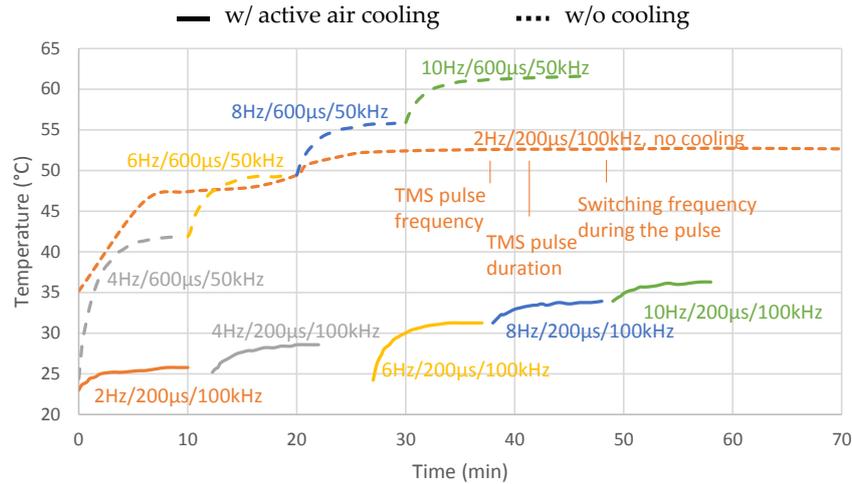

**Figure 11.** Temperature development of the SiC transistors under stress testing with sinusoidal current (700 A peak) per discrete transistor (enabling 8,400 A for a module with 12 transistors in our system) and transistor switching rates up to 100 kHz. Dashed lines were performed without any additional cooling in a closed 19" cabinet, solid lines included active air convection. The naming convention is pulse rate (Hz)/pulse duration /switching rate during the pulses.

Each module contains a complex programmable logic device (CPLD, MAX10, 10M02SCE144I7G, Altera/Intel, California, USA), which coordinates all fast processes such as switching, and an ARM-based microcontroller (ARM M4, MSP432P401RIPZR, Texas Instruments, Texas, USA, Arm, Cambridge, UK), which monitors the module, e.g., the temperature sensors. Both controllers can watch each other through hand-shake signals for proper operation and to identify failures. The overall control is performed on a central level by a system-on-chip controller (NI sbRIO 9627 with AMD/Xilinx Zynq 7000) and commands all module controllers through a bus system. The bus is isolated from the modules through signal isolators (ISO7841FDWW, Texas Instruments) and line drivers (UCC27712QDQ1, Texas Instruments).

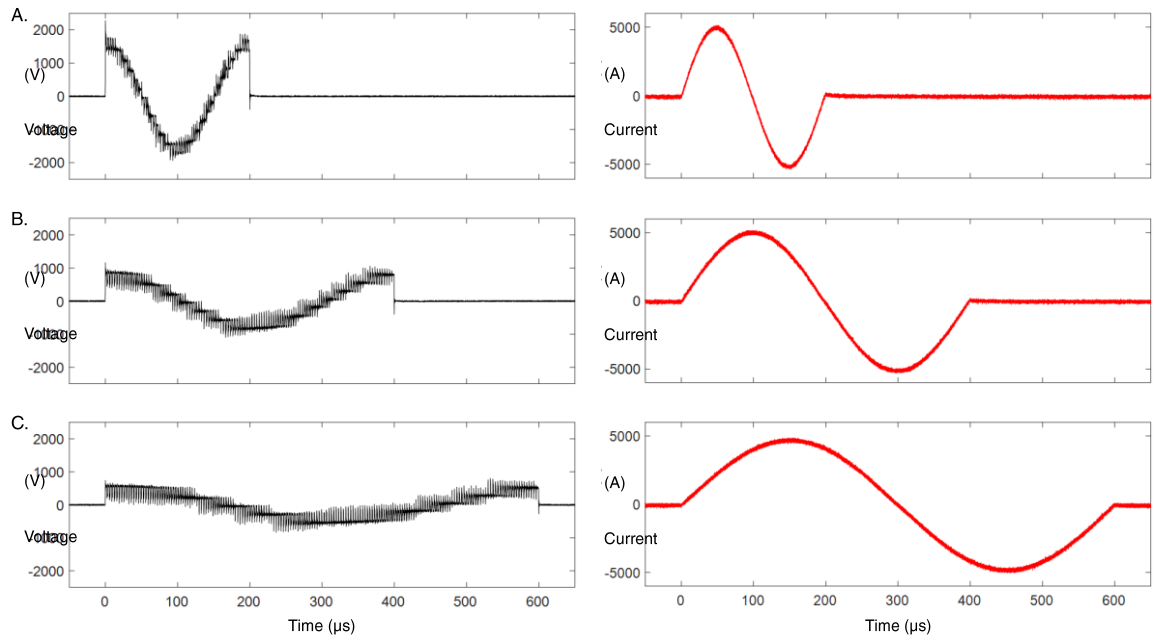

**Figure 12.** Recordings of biphasic pulses with cosinusoidal voltage as well as electric field shape and sinusoidal current shape. The pulse shape including its duration is purely software-generated (A 200 µs, B 400 µs, C 600 µs).

## Performance

### Switching and thermal performance

The major challenge of such is rapidly switching the high TMS pulse currents from one physical current path to another (commutation). As the current paths before and after a switching action cannot be perfectly identical, their magnetic fields and associated energy content differ. When these energies dissipate electrically in an uncontrolled manner, they lead to voltage spikes when the current is handed over between low-side and high-side transistor, which can damage the semiconductors. We performed double-pulse measurements through pumping an inductor (TMS coil D70 in this case, Magstim, Wales, UK) through the module as outlined in Figure 9 to form a triangular waveform and monitoring switching for all four conditions (commutation from high-side to low-side and from low-side to high-side for inwards and outwards current). The voltage was measured at the pins of the transistors and could be kept within the specified range up to transistor channel saturation. Modification of the transistor casing can reduce the parasitic inductance of the transistor itself and reduce voltage spikes for this application (Figure 10).

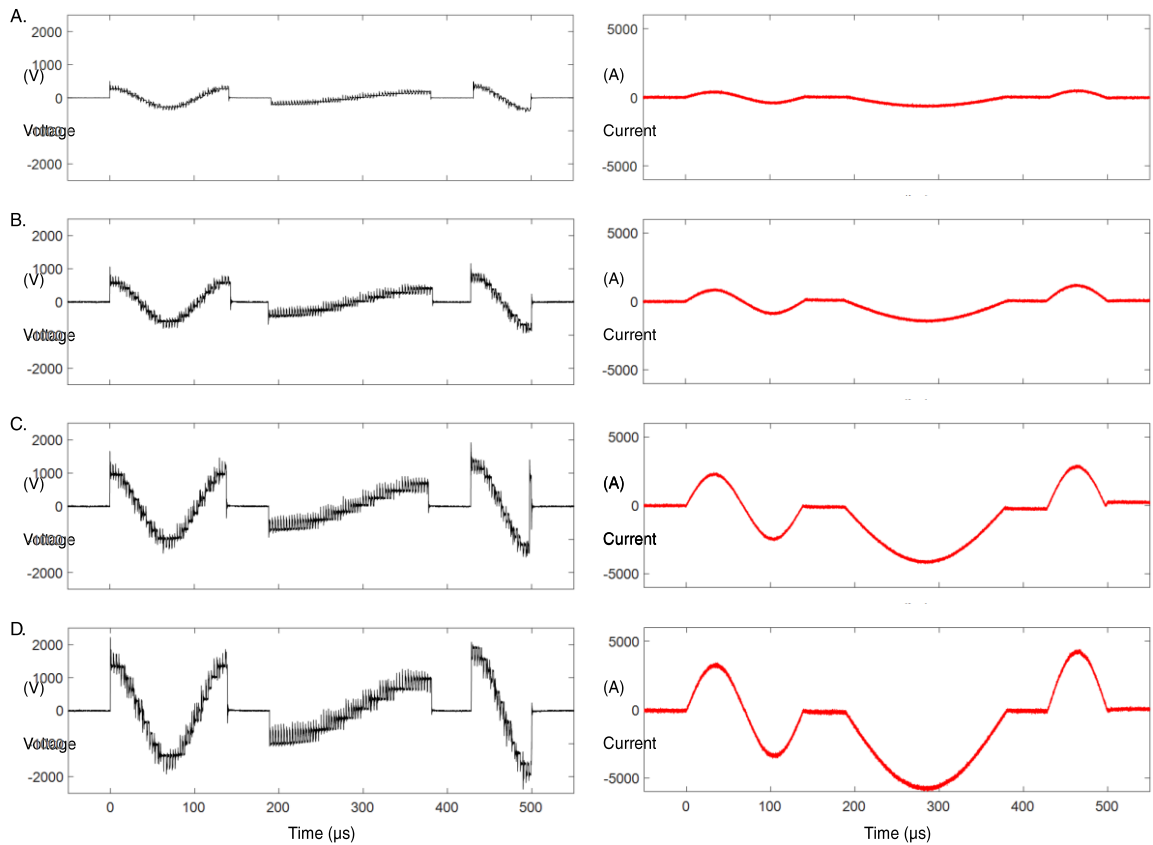

**Figure 13.** Recorded pulse train of typical conventional pulses (biphasic and two half-period pulses) with as little as 50 µs between two entirely different pulse shapes. The pulses have different energy content. Still, the device does not need any reconfiguration time or adjustment of charge to change the pulse duration, the pulse amplitude, or change the entire pulse shape. The panels from A to D show the same pulse sequence with increasing pulse amplitude. The recordings demonstrate both ways how the pulse synthesizer can adjust the pulse amplitude: Within a sequence, the circuit uses fewer steps for the lower-amplitude sections and pulses such as the long half-period pulse in the center. From pulse train to pulse train, the circuit increases the amplitude through increasing the module voltage step size.

Furthermore, the power electronics underwent stress testing with repetitive pulses to evaluate thermal limits under various switching conditions (see Figure 11). The test pulse current profile was bipolar sinusoidal with various pulse durations and peak levels of ± 700 A per discrete transistor. In each condition, the module performed approximately a million pulses to reach steady-state conditions of the (after all temperature-dependent) electrical and thermal conditions. The results in Figure 11 demonstrate that switching of up to 100 kHz per module is well within the performance range of the design so that six modules can safely reach an effective switching rate of 600 kHz.

### Synthesis of all conventional pulses with energy recovery

Figure 12 shows three biphasic pulses with sinusoidal current and cosinusoidal voltage shape with varying pulse durations. The electric-field pulse shape as the derivative of the current in the stimulation coil follows approximately the voltage shape if the inner resistance of the high-current loop formed by the electronics, cables, and the coil is low. The generation of the pulse shape is fully controlled throughout the pulse and does not depend on any hard-wired oscillator of a pulse-shaping circuit in conventional devices. Thus, the device can easily change the pulse duration as demonstrated in the figure. Compared to cTMS as the first TMS technology that could also rapidly change the pulse duration through software by actively terminating phases of an oscillator, the pulse does not have to be rectangular, but may be sinusoidal or have any other shape. The setup can provide sufficient voltage for short pulses, which need higher amplitude for stimulation, and energy for exceptionally long pulses with their large charge content per phase (Figure 12 C).

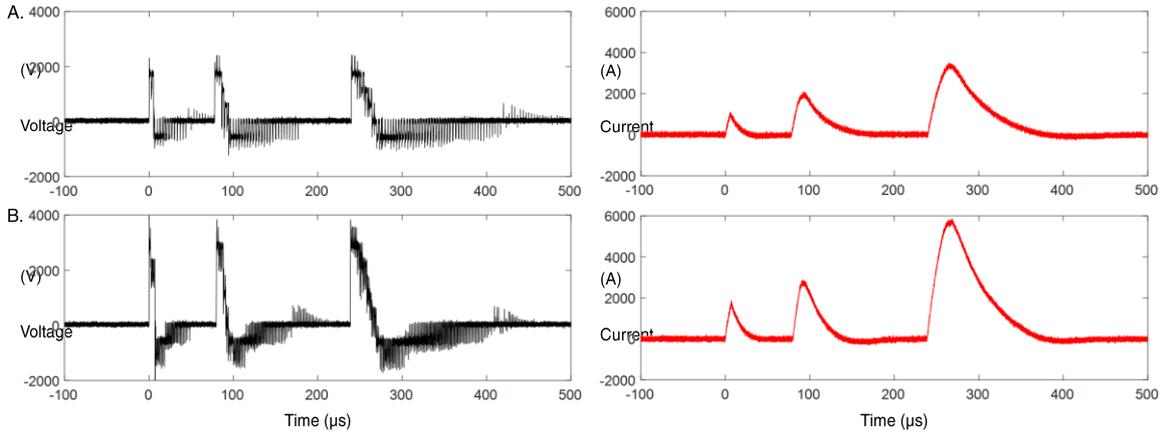

**Figure 14.** Sequence of monophasic pulses generated with various pulse durations and in two amplitudes. The approximately exponentially falling phase of the current in the second part of each pulse is generated through resistive damping in conventional monophasic TMS devices associated with the loss of the entire pulse energy. The pulse synthesizer, however, simulates only a resistor to the coil and extracts the energy back to the internal capacitors through proper control so that there is sufficient energy to generate such rapid sequences, which no conventional monophasic device can. Whereas we demonstrated the adjustment of the module voltages to tune the pulse strength in Figure 13, here the lower amplitude of the sequence in Panel A uses less module levels than Panel B to clarify the coarser quantization of this approach.

As a consequence of this software generation, the pulses also do not exhibit the typical amplitude reduction of some 20% of conventional biphasic TMS devices from the beginning of the pulse to the end, but the pulse voltage can end with the same amplitude as it started. In conventional devices, the amplitude reduction is a consequence of the energy loss in the oscillator; the conventional oscillator starts with a specific pulse energy and depletes throughout the pulse, resulting in decaying current and voltage amplitudes. The synthesizer does not have any such constraints so that the pulse can have constant amplitude over time, although it could imitate such damping.

Naturally, the device can generate other conventional pulse shapes, such as just half a period of a sinusoidal current and cosinusoidal voltage, sometimes called half-wave pulses (see

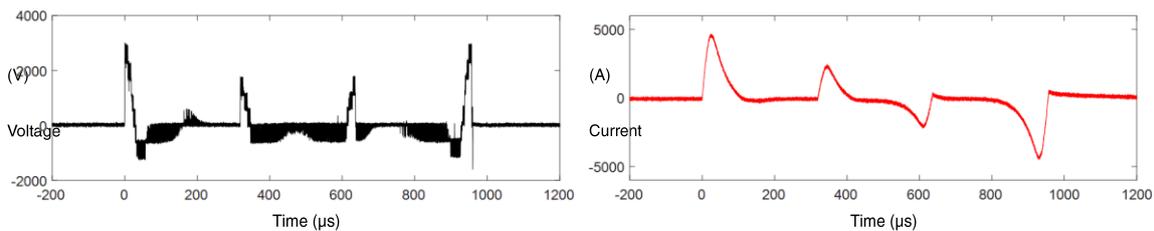

**Figure 15.** Burst of various monophasic pulses with different amplitude and time direction. To demonstrate the flexibility and ability to generate practically any pulse shape, the burst does not only change the amplitude but inverses the time direction for the third and fourth pulse so that the voltage is mirrored and the current additionally inverted.

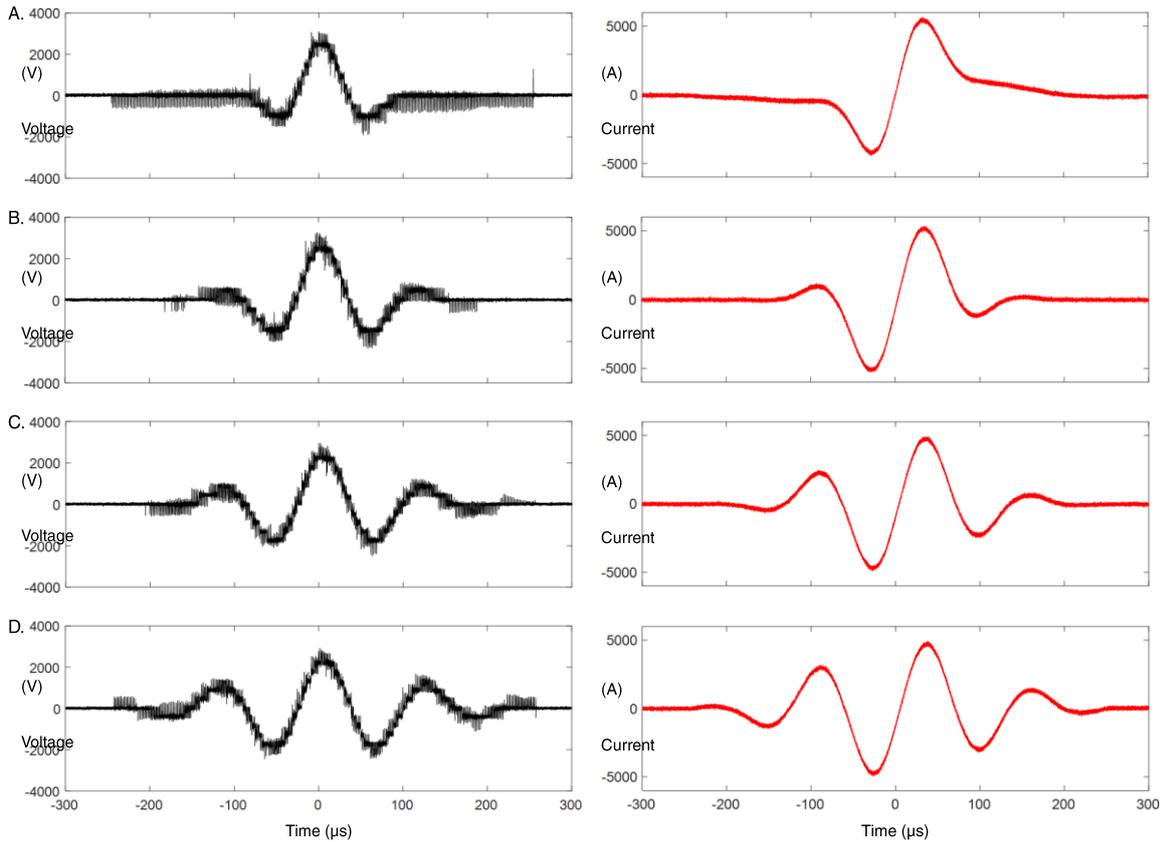

**Figure 16.** Polyphasic pulses with sinusoidal current and Gaussian envelope of increasing width from A to D. For very short envelope width, the pulse degenerates into a near-sinusoidal biphasic pulse, which in contrast to conventional biphasic TMS pulses contains smoother edges in the voltage.

Figure 13), and monophasic pulses (see Figure 14). As was already pointed out in our previous descriptions of the technology, the circuit does not contain any resistors for shaping a pulse [130, 131]. The abandonment of any resistance for pulse shaping, which is essential in conventional monophasic TMS devices and responsible for most of the problems of the pulse shape, has substantial impact: As a consequence of the energy recovery and the low energy loss even in monophasic pulses, the technology can generate monophasic pulses at repetition rates of 50 Hz and higher. As outlined below, the device can even synthesize monophasic pulses in rapid pulse trains with practically no time between pulses or mix it with other pulse shapes in a pulse train. The decay phase of the monophasic pulses, which conventional monophasic stimulators generate by connecting a powerful resistor to the coil and in which they convert the entire pulse energy into heat, is entirely simulated here only. The pulse synthesizer extracts the current of the coil so that the current decays with exponential shape and the voltage exhibits only a shallow low-amplitude counter-phase as intended. From the perspective of the coil, the machine thus still shows apparently resistive behavior, but the energy is widely recovered instead.

The device technology can use two ways to adjust the amplitude of a pulse. It can change the voltage of the modules, which changes the voltage step size and thus the granularity. In addition, the controller can modulate, i.e., calculate how to coordinate the modules to use the available voltage steps to generate the output at a different pulse amplitude. For lower amplitudes of the same pulse shape, it can, for instance, use fewer module steps. Fewer module steps, however, can decrease the output quality as the granularity becomes coarser and thus should be used only cautiously (see Figures 12 and 14).

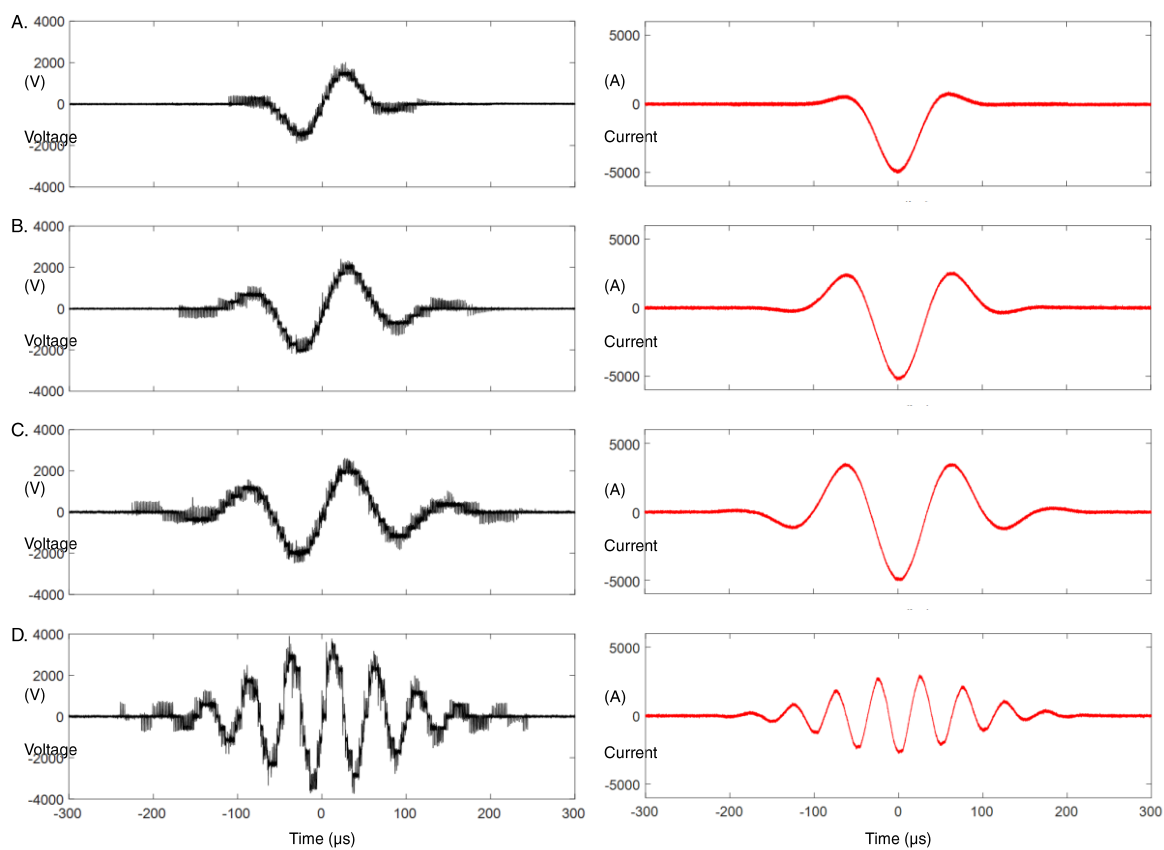

**Figure 17.** Polyphasic pulses with sinusoidal current and Gaussian envelope of increasing width from A to D. For very short envelope width, the pulse degenerates into a near-sinusoidal biphasic pulse, which in contrast to conventional biphasic TMS pulses contains smoother edges in the voltage.

### *Generation of novel and user-defined pulses*

Since pulse shaping is entirely a programmable control issue and performed by embedded software, the technology and its presented implementation can generate a wide range of pulse shapes with only few technological constraints, such as peak current, peak voltage, charge per phase without current reversal, and zero current at the end of the pulse. Importantly, due to the principle of the circuit, it generates all pulses without any exceptions with energy recovery, which previously was only known from conventional biphasic pulses.

Figure 15 demonstrates the ability of the device to deviate from the beaten track and generate unconventional pulses. The recording combines more conventional monophasic pulses with reversed ones, which no conventional TMS technology could possibly generate as the pulse now practically starts with a quasi-negative-resistive damping phase, which increases the pulse amplitude.

Polyphasic pulses, i.e., pulses with more than one sinusoidal current period were among the first in TMS and are promoted for special applications to this day due to their increased stimulation strength already at lower voltage and current amplitudes and under certain conditions presumably lower sound emission [91, 93, 114, 150, 151]. Whereas conventional technology, however, needs to use oscillator circuits, their amplitude is constant and rather falls during the pulse due to damping. Figure 16 shows pulses of sinusoidal currents with Gaussian envelope with varying width. Thus, the current oscillation builds up from an initially small amplitude below the noise floor to reach a maximum and subsequently decays again. For very short envelope widths, the pulse degenerates into a biphasic pulse with near-sinusoidal current but in contrast to conventional biphasic TMS pulses smoother leading and trailing edges in the voltage and the electric field. Neither the increasing amplitude during the first half nor the decay without damping is possible with conventional TMS circuitry or modifications of it. In Figure 17, the carrier wave under the Gaussian envelope is a cosine current. This pulse degenerates to a sinusoidal voltage for very short envelope widths.

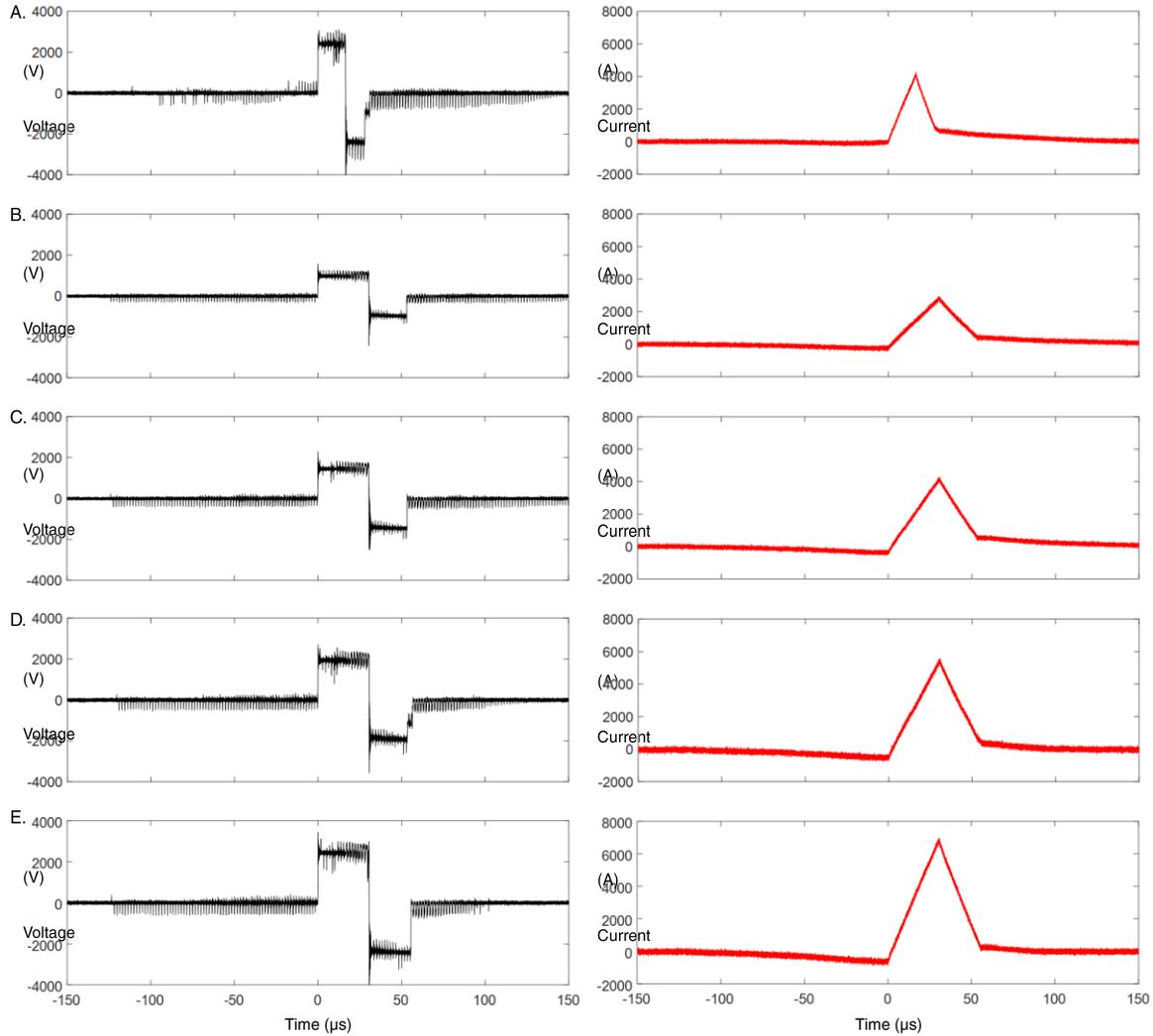

**Figure 18.** Pulse shapes optimized for minimum coil heating with different pulse duration (A and B) and various pulse amplitudes for the longer duration (B – E). Before the actual pulse, the device brings the current baseline slowly to a negative offset, from which the current can ramp up with a longer rising edge without reaching large currents. This pre-phase is generated through subtle control of the voltage in the approximately 120 µs introducing the pulse.

As another example of a pulse that conventional devices cannot generate, Figure 18 displays recordings of a loss-optimized biphasic pulse type that generates less coil heating than any standard pulse with adjusted stimulation strength as found previously [82, 83]. This pulse type includes a shallow pre-phase before the actual pulse, which has almost negligible electric field and therefore little influence on the neural activation. However, it generates an offset in the current baseline so that the current can subsequently contain a longer rising edge without reaching large current amplitudes. As the heating in the coil is mostly proportional to the squared current, the reduction of the peak over-compensates any loss of that pre-phase. Again, the already emphasized capability of the system to retrieve the pulse energy not only enables reducing coil heating with this pulse but furthermore also exploit the potential to reduce overall pulse losses in rapid trains.

## *Free combination of various pulses, pulse current directions, and pulse amplitudes in rapid succession*

The modular pulse synthesizer technology cannot only generate practically any TMS pulse, including all existing ones and a wide range of any user-defined pulses, but due to energy recovery also in rapid trains up to theta-burst and quadripulse stimulation as only little recharging is necessary. Beyond that, however, the technology can also change the pulse rapidly from pulse to pulse as needed for paired-pulse measurements with a weaker preconditioning pulse.

Figures 13 to 15 demonstrate that pulse trains can, for instance, combine pulses with different amplitude, current direction, pulse width, or entirely different shape. In the examples, the pulses are as close to each other as 50 µs to demonstrate the possible, though longer inter-pulse timings are unproblematic. Conventional devices, on the other hand, needed to physically reconfigure the circuit to switch, for example, from monophasic to biphasic by removing the damping resistor, which takes on the order of five to ten seconds in devices with such a feature. However, conventional device technology cannot even adjust the amplitude rapidly from pulse to pulse to a larger degree as amplitude adjustments requires charging or discharging the capacitor. The modular pulse synthesizer, however, can adjust the amplitude also through switching control as demonstrated above. The pulse shape including pulse duration and pulse direction are merely control aspects as outlined before and can be rapidly changed between pulses.

## *Conclusions*

We presented a modular pulse synthesizer TMS device with highly flexible pulse shape control and unparalleled quality of the output pulse. Neither conventional TMS nor conventional power electronic inverter circuits could achieve such pulse-shape control as the combination of power and bandwidth overwhelms transistors while standard inverter circuits furthermore generate poor electric field output quality with large high-frequency components. MPS TMS leveraged power electronics solutions we have developed over the past decade [130-133]. The technology combines a modular circuit that breaks the requisite high power and fast switching dynamics with latest unipolar wide-bandgap semiconductors to achieve an exceptionally high power–bandwidth product. The unipolar SiC field effect transistors do not exhibit the large and slow charge-storing effects of previous power semiconductor devices in that voltage range, enabling fast switching. The implemented device can cover the range of conventional TMS devices with up to ~ 4,000 V and ~ 8,000 A, with more than 600 kHz effective switching rate and fine output granularity to cover a bandwidth of at least 60 kHz with high quality. The modular design allows easy and practically unlimited scaling of all three performance figures of voltage, current, and effective switching rate.

The MPS TMS device can generate all conventional TMS pulse shapes and pulse protocols with low distortion and future novel pulse shapes with moderate constraints, such as user-defined or mathematically optimized pulses [82]. Notably, the modular pulse synthesizer principle generates pulses through electronic control, without any intentional resistance or damping. In consequence, all pulses, including monophasic ones, are generated with energy recycling [130, 131]. Conventional TMS circuits, in contrast, can recover some of the pulse energy only for biphasic pulses. Monophasic devices, in contrast, convert the entire pulse energy into heat due to their principles of pulse shaping and therefore only reach low repetition rates.

As a consequence of such energy recovery, MPS TMS can repeat all pulses, including monophasic ones, at high repetition rates.

More importantly, the technology can rapidly change any pulse features or the entire pulse shape. This rapid readjustment of pulses is in marked contrast to conventional machines, which typically could not even change the amplitude in larger steps or the current direction within rapid pulse trains or double pulses. Thus, MPS TMS can prime with one pulse amplitude or pulse shape with certain neural activation selectivity and follow up with another pulse with different amplitude and/or pulse shape with a different activation profile.

Beyond the high flexibility, the technology provides a number of technical advantages. In addition to the energy recovery for any pulse, we demonstrated the generation of pulses with lower coil heating for the same stimulation strength. Moreover, the circuit principle can generate pulse shapes differentially. Thus, in contrast to conventional TMS pulse sources, the average electrical potential of the coil is approximately constant to reduce the generation of capacitive artifacts in sensitive electronics, particularly neuroamplifiers, such as those for EMG, EEG, or intracranial recordings.

## *Acknowledgements*

Research reported in this publication was supported by the U.S. National Institutes of Health under award number RF1MH124943. The content is solely the responsibility of the authors and does not necessarily represent the official views of the National Institutes of Health. S. M. Goetz is inventor on patents and patent applications on MPS TMS, related power electronics circuits, and other TMS technology. Related to TMS technology, S. M. Goetz has previously received royalties from Rogue Research as well as research funding from Magstim; A. V. Peterchev is inventor on patents on TMS technology and has received research funding, travel support, patent royalties, consulting fees, equipment loans, hardware donations, and/or patent application support from Rogue Research, Magstim, MagVenture, Neuronetics, BTL Industries, and Advise Connect Inspire. Z. Li and J. Zhang declare no relevant conflict of interest.

# References

[1] P. M. Rossini, D. Burke, R. Chen, L. G. Cohen, Z. Daskalakis, R. Di Iorio*, et al.*, "Non-invasive electrical and magnetic stimulation of the brain, spinal cord, roots and peripheral nerves: Basic principles and procedures for routine clinical and research application. An updated report from an I.F.C.N. Committee," *Clinical Neurophysiology,* vol. 126, pp. 1071-1107, 2015.

[2] A. T. Barker, R. Jalinous, and I. L. Freeston, "Non-invasive magnetic stimulation of human motor cortex," *The Lancet,* vol. 325, pp. 1106-1107, 1985.

[3] Y. Ugawa, J. C. Rothwell, B. L. Day, P. D. Thompson, and C. D. Marsden, "Magnetic stimulation over the spinal enlargements," *Journal of Neurology, Neurosurgery & Psychiatry,* vol. 52, pp. 1025-1032, 1989.

[4] R. Nardone, Y. Höller, A. Taylor, A. Thomschewski, A. Orioli, V. Frey*, et al.*, "Noninvasive Spinal Cord Stimulation: Technical Aspects and Therapeutic Applications," *Neuromodulation: Technology at the Neural Interface,* vol. 18, pp. 580-591, 2015.

[5] M. J. R. Polson, A. T. Barker, and I. L. Freeston, "Stimulation of nerve trunks with time-varying magnetic fields," *Medical and Biological Engineering and Computing,* vol. 20, pp. 243-244, 1982.

[6] J. Szecsi, S. Goetz, W. Pöllmann, and A. Straube, "Force–pain relationship in functional magnetic and electrical stimulation of subjects with paresis and preserved sensation," *Clinical Neurophysiology,* vol. 121, pp. 1589-1597, 2010.

[7] S. M. Goetz, H. G. Herzog, N. Gattinger, and B. Gleich, "Comparison of coil designs for peripheral magnetic muscle stimulation," *Journal of Neural Engineering,* vol. 8, p. 056007, 2011.

[8] S. M. Goetz and Z.-D. Deng, "The development and modelling of devices and paradigms for transcranial magnetic stimulation," *International Review of Psychiatry,* vol. 29, pp. 115-145, 2017.

[9] L. J. Gomez, S. M. Goetz, and A. V. Peterchev, "Design of transcranial magnetic stimulation coils with optimal trade-off between depth, focality, and energy," *Journal of Neural Engineering,* vol. 15, p. 046033, 2018.

[10] A. Valero-Cabré, J. L. Amengual, C. Stengel, A. Pascual-Leone, and O. A. Coubard, "Transcranial magnetic stimulation in basic and clinical neuroscience: A comprehensive review of fundamental principles and novel insights," *Neuroscience & Biobehavioral Reviews,* vol. 83, pp. 381-404, 2017.

[11] S. Luan, I. Williams, K. Nikolic, and T. G. Constandinou, "Neuromodulation: present and emerging methods," *Frontiers in Neuroengineering,* vol. 7, 2014.

[12] A. T. Sack, R. C. Kadosh, T. Schuhmann, M. Moerel, V. Walsh, and R. Goebel, "Optimizing Functional Accuracy of TMS in Cognitive Studies: A Comparison of Methods," *Journal of Cognitive Neuroscience,* vol. 21, pp. 207-221, 2009.

[13] C. Miniussi, M. Ruzzoli, and V. Walsh, "The mechanism of transcranial magnetic stimulation in cognition," *Cortex,* vol. 46, pp. 128-130, 2010.

[14] B. Luber and S. H. Lisanby, "Enhancement of human cognitive performance using transcranial magnetic stimulation (TMS)," *NeuroImage,* vol. 85, pp. 961-970, 2014.

[15] Y. Levkovitz, M. Isserles, F. Padberg, S. H. Lisanby, A. Bystritsky, G. Xia*, et al.*, "Efficacy and safety of deep transcranial magnetic stimulation for major depression: a prospective multicenter randomized controlled trial," *World Psychiatry,* vol. 14, pp. 64-73, 2015.

[16] J. P. O'Reardon, H. B. Solvason, P. G. Janicak, S. Sampson, K. E. Isenberg, Z. Nahas*, et al.*, "Efficacy and Safety of Transcranial Magnetic Stimulation in the Acute Treatment of Major Depression: A Multisite Randomized Controlled Trial," *Biological Psychiatry,* vol. 62, pp. 1208-1216, 2007.

[17] M. S. George, S. H. Lisanby, D. Avery, W. M. McDonald, V. Durkalski, M. Pavlicova*, et al.*, "Daily Left Prefrontal Transcranial Magnetic Stimulation Therapy for Major Depressive Disorder: A Sham-Controlled Randomized TrialMagnetic Stimulation for Major Depressive Disorder," *Archives of General Psychiatry,* vol. 67, pp. 507-516, 2010.

[18] D. M. Blumberger, F. Vila-Rodriguez, K. E. Thorpe, K. Feffer, Y. Noda, P. Giacobbe*, et al.*, "Effectiveness of theta burst versus high-frequency repetitive transcranial magnetic stimulation in patients with depression (THREE-D): a randomised non-inferiority trial," *The Lancet,* vol. 391, pp. 1683-1692, 2018.

[19] U.S. Food and Drug Administration, "K122288; 510(K) Summary Brainsway Deep TMS System," www.accessdata.fda.gov/cdrh_docs/pdf12/K122288.pdf, 2013.

[20] A. Mantovani, H. B. Simpson, B. A. Fallon, S. Rossi, and S. H. Lisanby, "Randomized sham-controlled trial of repetitive transcranial magnetic stimulation in treatment-resistant obsessive–compulsive disorder," *International Journal of Neuropsychopharmacology,* vol. 13, pp. 217-227, 2010.

[21] R. B. Lipton, D. W. Dodick, S. D. Silberstein, J. R. Saper, S. K. Aurora, S. H. Pearlman*, et al.*, "Single-pulse transcranial magnetic stimulation for acute treatment of migraine with aura: a randomised, double-blind, parallel-group, sham-controlled trial," *The Lancet Neurology,* vol. 9, pp. 373-380, 2010.

[22] T. Picht, S. Schmidt, S. Brandt, D. Frey, H. Hannula, T. Neuvonen*, et al.*, "Preoperative Functional Mapping for Rolandic Brain Tumor Surgery: Comparison of Navigated Transcranial Magnetic Stimulation to Direct Cortical Stimulation," *Neurosurgery,* vol. 69, pp. 581-589, 2011.

[23] S. Takahashi and T. Picht, "Comparison of Navigated Transcranial Magnetic Stimulation to Direct Electrical Stimulation for Mapping the Motor Cortex Prior to Brain Tumor Resection," in *Tumors of the Central Nervous System, Volume 12: Molecular Mechanisms, Children's Cancer, Treatments, and Radiosurgery*, M. A. Hayat, Ed., ed Dordrecht: Springer Netherlands, 2014, pp. 261-276.

[24] A. Zangen, H. Moshe, D. Martinez, N. Barnea-Ygael, T. Vapnik, A. Bystritsky*, et al.*, "Repetitive transcranial magnetic stimulation for smoking cessation: a pivotal multicenter double-blind randomized controlled trial," *World Psychiatry,* vol. 20, pp. 397-404, 2021.

[25] M. C. Eldaief, D. Z. Press, and A. Pascual-Leone, "Transcranial magnetic stimulation in neurology: A review of established and prospective applications," *Neurology Clinical Practice,* vol. 3, pp. 519-526, 2013.

[26] T. Marzouk, S. Winkelbeiner, H. Azizi, A. K. Malhotra, and P. Homan, "Transcranial Magnetic Stimulation for Positive Symptoms in Schizophrenia: A Systematic Review," *Neuropsychobiology,* vol. 79, pp. 384-396, 2020.

[27] F. Fisicaro, G. Lanza, A. A. Grasso, G. Pennisi, R. Bella, W. Paulus*, et al.*, "Repetitive transcranial magnetic stimulation in stroke rehabilitation: review of the current evidence and pitfalls," *Therapeutic Advances in Neurological Disorders,* vol. 12, p. 1756286419878317, 2019.

[28] A. V. Peterchev, Z. D. Deng, and S. M. Goetz, "Advances in Transcranial Magnetic Stimulation Technology," in *Brain Stimulation*, ed: John Wiley & Sons, 2015, pp. 165-189.

[29] A. Thielscher and T. Kammer, "Electric field properties of two commercial figure-8 coils in TMS: calculation of focality and efficiency," *Clinical Neurophysiology,* vol. 115, pp. 1697-1708, 2004.

[30] Z.-D. Deng, S. H. Lisanby, and A. V. Peterchev, "Electric field depth–focality tradeoff in transcranial magnetic stimulation: Simulation comparison of 50 coil designs," *Brain Stimulation,* vol. 6, pp. 1-13, 2013.

[31] A. Karabanov, U. Ziemann, M. Hamada, M. S. George, A. Quartarone, J. Classen*, et al.*, "Consensus Paper: Probing Homeostatic Plasticity of Human Cortex With Non-invasive Transcranial Brain Stimulation," *Brain Stimulation,* vol. 8, pp. 993-1006, 2015.

[32] M. G. Stokes, A. T. Barker, M. Dervinis, F. Verbruggen, L. Maizey, R. C. Adams*, et al.*, "Biophysical determinants of transcranial magnetic stimulation: effects of excitability and depth of targeted area," *Journal of Neurophysiology,* vol. 109, pp. 437-444, 2012.

[33] A. S. Aberra, B. Wang, W. M. Grill, and A. V. Peterchev, "Simulation of transcranial magnetic stimulation in head model with morphologically-realistic cortical neurons," *Brain Stimulation,* vol. 13, pp. 175-189, 2020.

[34] E. R. Lontis, M. Voigt, and J. J. Struijk, "Focality Assessment in Transcranial Magnetic Stimulation With Double and Cone Coils," *Journal of Clinical Neurophysiology,* vol. 23, pp. 463-472, 2006.

[35] C. I. Bargmann and W. T. Newsome, "The Brain Research Through Advancing Innovative Neuro-technologies (BRAIN) Initiative and Neurology," *JAMA Neurology,* vol. 71, pp. 675-676, 2014.

[36] E. M. Mosier, M. Wolfson, E. Ross, J. Harris, D. Weber, and K. A. Ludwig, "Chapter 5: The Brain Initiative—Implications for a Revolutionary Change in Clinical Medicine via Neuromodulation Technology," in *Neuromodulation (Second Edition)*, E. S. Krames, P. H. Peckham, and A. R. Rezai, Eds., ed: Academic Press, 2018, pp. 55-68.

[37] V. Di Lazzaro, J. Rothwell, and M. Capogna, "Noninvasive Stimulation of the Human Brain: Activation of Multiple Cortical Circuits," *The Neuroscientist,* vol. 24, pp. 246-260, 2017.

[38] G. Strigaro, M. Hamada, C. R, and J. C. Rothwell, "Variability in response to 1 Hz repetitive TMS," *Brain Stimulation: Basic, Translational, and Clinical Research in Neuromodulation,* vol. 8, pp. 383-384, 2015.

[39] F. Maeda, J. P. Keenan, J. M. Tormos, H. Topka, and A. Pascual-Leone, "Interindividual variability of the modulatory effects of repetitive transcranial magnetic stimulation on cortical excitability," *Experimental Brain Research,* vol. 133, pp. 425-430, 2000.

[40] A. Guerra, V. López-Alonso, B. Cheeran, and A. Suppa, "Variability in non-invasive brain stimulation studies: Reasons and results," *Neuroscience Letters,* 2017.

[41] Y.-Z. Huang, M.-K. Lu, A. Antal, J. Classen, M. Nitsche, U. Ziemann*, et al.*, "Plasticity induced by non-invasive transcranial brain stimulation: A position paper," *Clinical Neurophysiology,* vol. 128, pp. 2318-2329, 2017.

[42] S. M. Goetz, B. Luber, S. H. Lisanby, D. L. Murphy, I. C. Kozyrkov, W. M. Grill*, et al.*, "Enhancement of Neuromodulation with Novel Pulse Shapes Generated by Controllable Pulse Parameter Transcranial Magnetic Stimulation," *Brain Stimul,* vol. 9, pp. 39-47, 2016.

[43] J. Rothwell and R. Hannah, "Effects of pulse width on responses to single, double and repetitive TMS of motor cortex," *Brain Stimulation,* vol. 12, p. 588, 2019.

[44] K. D'Ostilio, S. M. Goetz, M. Ciocca, R. Chieffo, J.-C. A. Chen, A. V. Peterchev*, et al.*, "Effect of coil orientation on strength-duration time constant with controllable pulse parameter transcranial magnetic stimulation," *Clinical Neurophysiology,* vol. 125 (Suppl. 1), p. S123, 2014.

[45] E. P. Casula, L. Rocchi, R. Hannah, and J. C. Rothwell, "Effects of pulse width, waveform and current direction in the cortex: A combined cTMS-EEG study," *Brain Stimulation,* vol. 11, pp. 1063-1070, 2018.

[46] M. Sommer, M. Ciocca, R. Chieffo, P. Hammond, A. Neef, W. Paulus*, et al.*, "TMS of primary motor cortex with a biphasic pulse activates two independent sets of excitable neurones," *Brain Stimulation,* vol. 11, pp. 558-565, 2018.

[47] R. Hannah and J. C. Rothwell, "Pulse Duration as Well as Current Direction Determines the Specificity of Transcranial Magnetic Stimulation of Motor Cortex during Contraction," *Brain Stimulation,* vol. 10, pp. 106-115, 2017.

[48] Y. Shirota, S. Dhaka, W. Paulus, and M. Sommer, "Current direction-dependent modulation of human hand motor function by intermittent theta burst stimulation (iTBS)," *Neuroscience Letters,* vol. 650, pp. 109-113, 2017.

[49] M. Sommer, A. Alfaro, M. Rummel, S. Speck, N. Lang, T. Tings*, et al.*, "Half sine, monophasic and biphasic transcranial magnetic stimulation of the human motor cortex," *Clinical Neurophysiology,* vol. 117, pp. 838-844, 2006.

[50] M. Sommer, C. Norden, L. Schmack, H. Rothkegel, N. Lang, and W. Paulus, "Opposite Optimal Current Flow Directions for Induction of Neuroplasticity and Excitation Threshold in the Human Motor Cortex," *Brain Stimulation,* vol. 6, pp. 363-370, 2013.

[51] T. Kammer, S. Beck, M. Erb, and W. Grodd, "The influence of current direction on phosphene thresholds evoked by transcranial magnetic stimulation," *Clinical Neurophysiology,* vol. 112, pp. 2015-2021, 2001.

[52] T. Kammer, S. Beck, A. Thielscher, U. Laubis-Herrmann, and H. Topka, "Motor thresholds in humans: a transcranial magnetic stimulation study comparing different pulse waveforms, current directions and stimulator types," *Clinical Neurophysiology,* vol. 112, pp. 250-258, 2001.

[53] T. Kammer, M. Vorwerg, and B. Herrnberger, "Anisotropy in the visual cortex investigated by neuronavigated transcranial magnetic stimulation," *NeuroImage,* vol. 36, pp. 313-321, 2007.

[54] V. Lazzaro, A. Oliviero, P. Mazzone, A. Insola, F. Pilato, E. Saturno*, et al.*, "Comparison of descending volleys evoked by monophasic and biphasic magnetic stimulation of the motor cortex in conscious humans," *Experimental Brain Research,* vol. 141, pp. 121-127, 2001.

[55] D. Spampinato, "Dissecting two distinct interneuronal networks in M1 with transcranial magnetic stimulation," *Experimental Brain Research,* 2020.

[56] R. Hannah, L. Rocchi, S. Tremblay, E. Wilson, and J. C. Rothwell, "Pulse width biases the balance of excitation and inhibition recruited by transcranial magnetic stimulation," *Brain Stimulation: Basic, Translational, and Clinical Research in Neuromodulation,* vol. 13, pp. 536-538, 2020.

[57] P. Davila-Pérez, A. Jannati, P. J. Fried, J. Cudeiro Mazaira, and A. Pascual-Leone, "The Effects of Waveform and Current Direction on the Efficacy and Test–Retest Reliability of Transcranial Magnetic Stimulation," *Neuroscience,* vol. 393, pp. 97-109, 2018.

[58] I. Halawa, Y. Shirota, A. Neef, M. Sommer, and W. Paulus, "Neuronal tuning: Selective targeting of neuronal populations via manipulation of pulse width and directionality," *Brain Stimulation,* vol. 12, p. 399, 2019.

[59] I. Halawa, Y. Shirota, A. Neef, M. Sommer, and W. Paulus, "Longer cTMS pulse width switches 1 Hz inhibitory motor cortex rTMS aftereffects to excitation," *Brain Stimulation: Basic, Translational, and Clinical Research in Neuromodulation,* vol. 12, p. 589, 2019.

[60] M. Sommer, N. Lang, F. Tergau, and W. Paulus, "Neuronal tissue polarization induced by repetitive transcranial magnetic stimulation?," *Neuroreport,* vol. 13, pp. 809-11, 2002.

[61] A. Antal, T. Z. Kincses, M. A. Nitsche, O. Bartfai, I. Demmer, M. Sommer*, et al.*, "Pulse configuration-dependent effects of repetitive transcranial magnetic stimulation on visual perception," *Neuroreport,* vol. 13, pp. 2229-33, 2002.

[62] T. Tings, N. Lang, F. Tergau, W. Paulus, and M. Sommer, "Orientation-specific fast rTMS maximizes corticospinal inhibition and facilitation," *Exp Brain Res,* vol. 164, pp. 323-33, 2005.

[63] K. Nakamura, S. J. Groiss, M. Hamada, H. Enomoto, S. Kadowaki, M. Abe*, et al.*, "Variability in Response to Quadripulse Stimulation of the Motor Cortex," *Brain Stimulation,* vol. 9, pp. 859-866, 2016.

[64] J. L. Taylor and C. K. Loo, "Stimulus waveform influences the efficacy of repetitive transcranial magnetic stimulation," *Journal of Affective Disorders,* vol. 97, pp. 271-276, 2007.

[65] Y. Hosono, R. Urushihara, M. Harada, N. Morita, N. Murase, Y. Kunikane*, et al.*, "Comparison of monophasic versus biphasic stimulation in rTMS over premotor cortex: SEP and SPECT studies," *Clinical Neurophysiology,* vol. 119, pp. 2538-2545, 2008.

[66] D. McRobbie, "Design and instrumentation of a magnetic nerve stimulator," *Journal of Physics E: Scientific Instruments,* vol. 18, p. 74, 1985.

[67] M. Schmid, T. Weyh, and B.-U. Meyer, "Entwicklung, Optimierung und Erprobung neuer Geräte für die magnetomotorische Stimulation von Nervenfasern / Development, Optimization, and Testing of New Devices for Magnetomotive Fibre Stimulation," *Biomedizinische Technik / Biomedical Engineering,* vol. 38, pp. 317-324, 1993. doi: 10.1515/bmte.1993.38.12.317

[68] Y.-Z. Huang, M. J. Edwards, E. Rounis, K. P. Bhatia, and J. C. Rothwell, "Theta Burst Stimulation of the Human Motor Cortex," *Neuron,* vol. 45, pp. 201-206, 2005.

[69] N. H. Jung, B. Gleich, N. Gattinger, C. Hoess, C. Haug, H. R. Siebner*, et al.*, "Quadri-Pulse Theta Burst Stimulation using Ultra-High Frequency Bursts – A New Protocol to Induce Changes in Cortico-Spinal Excitability in Human Motor Cortex," *PLOS ONE,* vol. 11, p. e0168410, 2016.

[70] M. Hamada and Y. Ugawa, "Quadripulse stimulation – A new patterned rTMS," *Restorative Neurology and Neuroscience,* vol. 28, pp. 419-424, 2010.

[71] S. Simeoni, R. Hannah, D. Sato, M. Kawakami, J. Rothwell, S. Simeoni*, et al.*, "Effects of Quadripulse Stimulation on Human Motor Cortex Excitability: A Replication Study," *Brain Stimulation,* vol. 9, pp. 148-150, 2016.

[72] K. D'Ostilio, S. M. Goetz, R. Hannah, M. Ciocca, R. Chieffo, J.-C. A. Chen*, et al.*, "Effect of coil orientation on strength–duration time constant and I-wave activation with controllable pulse parameter transcranial magnetic stimulation," *Clinical Neurophysiology,* vol. 127, pp. 675-683, 2016.

[73] M. Sommer, K. D'Ostilio, M. Cioccia, R. Hannah, P. Hammond, S. Goetz*, et al.*, "TMS can selectively activate and condition two different sets of excitatory synaptic inputs to corticospinal neurons in humans," *SFN Society for Neuroscience,* p. 542, 2014.

[74] A. V. Peterchev, B. Luber, G. G. Westin, and S. H. Lisanby, "Pulse Width Affects Scalp Sensation of Transcranial Magnetic Stimulation," *Brain Stimulation,* vol. 10, pp. 99-105, 2017.

[75] S. M. Goetz, D. L. K. Murphy, and A. V. Peterchev, "Transcranial Magnetic Stimulation Device With Reduced Acoustic Noise," *IEEE Magnetics Letters,* vol. 5, pp. 1-4, 2014.

[76] S. M. Goetz, S. H. Lisanby, D. L. K. Murphy, R. J. Price, G. O'Grady, and A. V. Peterchev, "Impulse Noise of Transcranial Magnetic Stimulation: Measurement, Safety, and Auditory Neuromodulation," *Brain Stimulation: Basic, Translational, and Clinical Research in Neuromodulation,* vol. 8, pp. 161-163, 2015.

[77] A. V. Peterchev, D. L. K. Murphy, and S. M. Goetz, "Quiet transcranial magnetic stimulation: Status and future directions," in *2015 37th Annual International IEEE Proc. Engin. Med. Biol. (EMBC)*, 2015, vol. 37, pp. 226-229. doi 10.1109/EMBC.2015.7318341

[78] Z. Zeng, L. M. Koponen, R. Hamdan, Z. Li, S. M. Goetz, and A. V. Peterchev, "Modular multilevel TMS device with wide output range and ultrabrief pulse capability for sound reduction," *bioRxiv,* 2021.09.08.459501, 2021.

[79] L. M. Koponen, S. M. Goetz, and A. V. Peterchev, "Double-Containment Coil With Enhanced Winding Mounting for Transcranial Magnetic Stimulation With Reduced Acoustic Noise," *IEEE Transactions on Biomedical Engineering,* vol. 68, pp. 2233-2240, 2021.

[80] S. A. Counter, "Auditory brainstem and cortical responses following extensive transcranial magnetic stimulation," *Journal of the Neurological Sciences,* vol. 124, pp. 163-170, 1994.

[81] W. C. Clapp, J. P. Hamm, I. J. Kirk, and T. J. Teyler, "Translating Long-Term Potentiation from Animals to Humans: A Novel Method for Noninvasive Assessment of Cortical Plasticity," *Biological Psychiatry,* vol. 71, pp. 496-502, 2012.

[82] S. M. Goetz, C. N. Truong, M. G. Gerhofer, A. V. Peterchev, H.-G. Herzog, and T. Weyh, "Analysis and Optimization of Pulse Dynamics for Magnetic Stimulation," *PLoS ONE,* vol. 8, p. e55771, 2013.

[83] S. M. Goetz, N. C. Truong, M. G. Gerhofer, A. V. Peterchev, H. G. Herzog, and T. Weyh, "Optimization of magnetic neurostimulation waveforms for minimum power loss," *IEEE Proc. Engin. Med. Biol.,* vol. 34, pp. 4652-4655, 2012. doi: 10.1109/EMBC.2012.6347004

[84] S. Rossi, M. Hallett, P. M. Rossini, and A. Pascual-Leone, "Safety, ethical considerations, and application guidelines for the use of transcranial magnetic stimulation in clinical practice and research," *Clinical Neurophysiology,* vol. 120, pp. 2008-2039, 2009.

[85] *UL 60601-1 Medical Electrical Equipment, Part 1: General Requirements for Safety,* derived from IEC 60601-1, Underwriters Laboratories, 2012.

[86] A. V. Peterchev, S. M. Goetz, G. G. Westin, B. Luber, and S. H. Lisanby, "Pulse width dependence of motor threshold and input–output curve characterized with controllable pulse parameter transcranial magnetic stimulation," *Clinical Neurophysiology,* vol. 124, pp. 1364-1372, 2013.

[87] A. T. Barker, C. W. Garnham, and I. L. Freeston, "Magnetic nerve stimulation: the effect of waveform on efficiency, determination of neural membrane time constants and the measurement of stimulator output," *Electroencephalogr Clin Neurophysiol Supplement,* vol. 43, pp. 227-37, 1991.

[88] R. Jalinous, "Principles of magnetic stimulator design," in *Handbook of Transcranial Magnetic Stimulation*, A. Pascual-Leone, N. J. Davey, J. C. Rothwell, E. M. Wassermann, and B. K. Puri, Eds., ed London: Arnold, 2002.

[89] H. Rothkegel, M. Sommer, W. Paulus, and N. Lang, "Impact of pulse duration in single pulse TMS," *Clinical Neurophysiology,* vol. 121, pp. 1915-1921, 2010.

[90] A. V. Peterchev, Z.-D. Deng, and S. M. Goetz, "Advances in transcranial magnetic stimulation technology," in *Brain Stimulation: Methodologies and Interventions*, I. M. Reti, Ed., ed Hoboken, NJ: John Wiley & Sons, 2015.

[91] D. Claus, N. M. F. Murray, A. Spitzer, and D. Flügel, "The influence of stimulus type on the magnetic excitation of nerve structures," *Electroencephalography and Clinical Neurophysiology,* vol. 75, pp. 342-349, 1990.

[92] R. Jalinous, "Technical and Practival Aspects of Magnetic Nerve Stimulation," *Journal of Clinical Neurophysiology,* vol. 8, pp. 10-25, 1991.

[93] J. Cadwell, "Optimizing magnetic stimulator design," *Electroencephalography and Clinical Neurophysiology. Supplement,* vol. 43, pp. 238-248, 1991.

[94] M. J. Wessel, L. R. Draaisma, T. Morishita, and F. C. Hummel, "The Effects of Stimulator, Waveform, and Current Direction on Intracortical Inhibition and Facilitation: A TMS Comparison Study," *Frontiers in Neuroscience,* vol. 13, 2019.

[95] A. V. Peterchev, K. D'Ostilio, J. C. Rothwell, and D. L. Murphy, "Controllable pulse parameter transcranial magnetic stimulator with enhanced circuit topology and pulse shaping," *Journal of Neural Engineering,* vol. 11, p. 056023, 2014.

[96] A. T. Barker, C. W. Garnham, and I. L. Freeston, "Magnetic nerve stimulation: the effect of waveform on efficiency, determination of neural membrane time constants and the measurement of stimulator output," *Electroencephalography and Clinical Neurophysiology. Supplement,* vol. 43, pp. 227-237, 1991.

[97] N. Lang, J. Harms, T. Weyh, R. N. Lemon, W. Paulus, J. C. Rothwell*, et al.*, "Stimulus intensity and coil characteristics influence the efficacy of rTMS to suppress cortical excitability," *Clinical Neurophysiology,* vol. 117, pp. 2292-2301, 2006.

[98] A. V. Peterchev, D. L. Murphy, and S. H. Lisanby, "Repetitive transcranial magnetic stimulator with controllable pulse parameters," *Journal of Neural Engineering,* vol. 8, p. 036016, 2011.

[99] A. V. Peterchev, R. Jalinous, and S. H. Lisanby, "A Transcranial Magnetic Stimulator Inducing Near-Rectangular Pulses With Controllable Pulse Width (cTMS)," *IEEE Transactions on Biomedical Engineering,* vol. 55, pp. 257-266, 2008.

[100] M. Moritz, F. Schmitt, P. Schweighofer, and P. Havel, "Magnetic Stimulation Device," Siemens AG, US 6,450,940; EP 0 996 485; European Patent Office, U.S. Patent and Trademark Office 2002.

[101] N. Gattinger, G. Mossnang, and B. Gleich, "flexTMS – A Novel Repetitive Transcranial Magnetic Stimulation Device With Freely Programmable Stimulus Currents," *Biomedical Engineering, IEEE Transactions on,* vol. 59, pp. 1962-1970, 2012.

[102] L. M. Koponen, J. O. Nieminen, T. P. Mutanen, M. Stenroos, and R. J. Ilmoniemi, "Coil optimisation for transcranial magnetic stimulation in realistic head geometry," *Brain Stimulation,* vol. 10, pp. 795-805, 2017.

[103] A. V. Peterchev, "Circuit topology comparison and design analysis for controllable pulse parameter transcranial magnetic stimulators," in *Neural Engineering (NER), 2011 5th International IEEE/EMBS Conference on*, 2011, pp. 646-649.

[104] E. S. Boyden, M. Kim, G. Abram, and M. Henninger, "Portable modular transcranial magnetic stimulation device," Massachusetts Institute of Technology, US 2009/0018384, US 12/117,896, U.S. Patent and Trademark Office, 2009.

[105] M. Talebinejad and A. D. C. Chan, "Circuit and method for use in transranial magnetic stimulation," US 9,504,846, US 14/442,543, U.S. Patent and Trademark Office, 2016.

[106] Y. Roth, A. Zangen, V. Chudnovsky, N. Safra, and D. Hazani, "Systems and Methods for Controlling Electric Field Pulse Parameters Using Transcranial Magnetic Stimulation," Yeda Research & Development Co. Ltd. at the Weizmann Institute of Science, US 9,180,305; EP 2 376 178 B1; US 2010/0152522, European Patent Office, U.S. Patent and Trademark Office, 2015.

[107] S. M. Goetz and T. Weyh, "Vorrichtung zur Nervenreizung mit Magnetfeldimpulsen," DE 10 2009 023 855 B4, DE 10 2009 023 855.7, Deutsches Patent- und Markenamt (DPMA), 2009.

[108] E. Corthout, A. Barker, and A. Cowey, "Transcranial magnetic stimulation. Which part of the current waveform causes the stimulation?," *Experimental Brain Research,* vol. 141, pp. 128-132, 2001.

[109] M. Riehl, "TMS stimulator design," in *Oxford Handbook of Transcranial Stimulation*, E. M. Wassermann, C. M. Epstein, U. Ziemann, V. Walsh, T. Paus, and S. H. Lisanby, Eds., ed Oxford: Oxford University Press, 2008, pp. 13-23.

[110] C. M. Epstein, "A six-pound battery-powered portable transcranial magnetic stimulator," *Brain Stimulation,* vol. 1, pp. 128-130, 2008.

[111] S. Machida, K. Ito, and Y. Yamashita, "Approaching the limit of switching loss reduction in Si-IGBTs," in *Power Semiconductor Devices & IC's (ISPSD), 2014 IEEE 26th International Symposium on*, 2014, pp. 107-110. doi: 10.1109/ISPSD.2014.6855987

[112] J. Qian, A. Khan, and I. Batarseh, "Turn-off switching loss model and analysis of IGBT under different switching operation modes," in *IEEE Proceedings of the Industrial Electronics, Control (IECON),* 1995, vol. 21, pp. 240-245. doi: 10.1109/IECON.1995.483365

[113] A. C. Oliveira, C. B. Jacobina, and A. M. N. Lima, "Improved Dead-Time Compensation for Sinusoidal PWM Inverters Operating at High Switching Frequencies," *Industrial Electronics, IEEE Transactions on,* vol. 54, pp. 2295-2304, 2007.

[114] D. Emrich, A. Fischer, C. Altenhöfer, T. Weyh, F. Helling, S. Goetz*,* M. Brielmeier, and K. Matiasek, "Muscle force development after low-frequency magnetic burst stimulation in dogs," *Muscle & Nerve,* vol. 46, pp. 951-953, 2012.

[115] A. M. Hermsen, A. Haag, C. Duddek, K. Balkenhol, H. Bugiel, S. Bauer, *et al.*, "Test–retest reliability of single and paired pulse transcranial magnetic stimulation parameters in healthy subjects," *Journal of the Neurological Sciences,* vol. 362, pp. 209-216, 2016.

[116] F. Ferreri, P. Pasqualetti, S. Määttä, D. Ponzo, F. Ferrarelli, G. Tononi, *et al.*, "Human brain connectivity during single and paired pulse transcranial magnetic stimulation," *NeuroImage,* vol. 54, pp. 90-102, 2011.

[117] B. Boroojerdi, L. Kopylev, F. Battaglia, S. Facchini, U. Ziemann, W. Muellbacher, *et al.*, "Reproducibility of intracortical inhibition and facilitation using the paired-pulse paradigm," *Muscle & Nerve,* vol. 23, pp. 1594-1597, 2000.

[118] J. Reis, O. B. Swayne, Y. Vandermeeren, M. Camus, M. A. Dimyan, M. Harris-Love, *et al.*, "Contribution of transcranial magnetic stimulation to the understanding of cortical mechanisms involved in motor control," *The Journal of Physiology,* vol. 586, pp. 325-351, 2008.

[119] P. Schweighofer, M. Moritz, and F. Schmitt, "Magnetic stimulation device," Siemens AG, US 6,123,658; EP 0 958 844, US 09/313,559, European Patent Office, U.S. Patent and Trademark Office, 2000.

[120] A. Emadi, L. Young Joo, and K. Rajashekara, "Power Electronics and Motor Drives in Electric, Hybrid Electric, and Plug-In Hybrid Electric Vehicles," *Industrial Electronics, IEEE Transactions on,* vol. 55, pp. 2237-2245, 2008.

[121] S. Goetz and C. Korte, "Spectral synthesis of switching distortion in automotive drive inverters," in *Proceedings of the European IEEE Conference on Power Electronics and Applications (EPE'16 ECCE Europe)*, vol. 18, 2016. doi: 10.1109/EPE.2016.7695632

[122] M. J. R. Polson, "Magnetic stimulator with increased energy efficiency," Magstim Co., Patent, GB 2 298 370, UK Intellectual Property Office, 1996.

[123] M. J. R. Polson, "Magnetic stimulator for neuro-muscular tissue," Magstim Co., U.S. Patent and Trademark Office, 1998.

[124] K. Mazac, "Verfahren und Vorrichtung zur Anwendung von elektromagnetischen Feldern (EMF) in therapeutischer Praxis," DE 10 2010 009 743, DE 10 2010 009 743, Deutsches Patent- und Markenamt (DPMA), 2010.

[125] D. J. Perreault and S. Mogren, "Magnetic stimulator power and control circuit," R. B. Carr Engineering, US 6,551,233 B2, US 09/924,907, U.S. Patent and Trademark Office, 2003.

[126] M. A. Moffitt, C. C. Mcintyre, and W. M. Grill, "Prediction of myelinated nerve fiber stimulation thresholds: limitations of linear models," *IEEE Transactions on Biomedical Engineering,* vol. 51, pp. 229-236, 2004.

[127] H. Motz and F. Rattay, "A study of the application of the Hodgkin-Huxley and the Frankenhaeuser-Huxley model for electrostimulation of the acoustic nerve," *Neuroscience,* vol. 18, pp. 699-712, 1986.

[128] E. M. Izhikevich, "Resonate-and-fire neurons," *Neural Networks,* vol. 14, pp. 883-894, 2001.

[129] C. C. McIntyre, A. G. Richardson, and W. M. Grill, "Modeling the Excitability of Mammalian Nerve Fibers: Influence of Afterpotentials on the Recovery Cycle," *Journal of Neurophysiology,* vol. 87, pp. 995-1006, 2002.

[130] S. M. Goetz, M. Pfaeffl, J. Huber, M. Singer, R. Marquardt, and T. Weyh, "Circuit topology and control principle for a first magnetic stimulator with fully controllable waveform," *Proc. IEEE Eng. Med. Biol. EMBC,* vol. 34, pp. 4700-4703, 2012. doi: 10.1109/EMBC.2012.6347016

[131] S. M. Goetz, M. Singer, J. Huber, M. Pfaeffl, R. Marquardt, and T. Weyh, "Magnetic Stimulation with Arbitrary Waveform Shapes," in *IFMBE Proceedings, Medical Physics and Biomedical Engineering*, Beijing, 2013, pp. 2244-2247. doi: 10.1007/978-3-642-29305-4_589

[132] S. M. Goetz, A. V. Peterchev, and T. Weyh, "Modular Multilevel Converter With Series and Parallel Module Connectivity: Topology and Control," *Power Electronics, IEEE Transactions on,* vol. 30, pp. 203-215, 2015.

[133] S. Goetz and T. Weyh, "Magnetic stimulation having a freely selectable pulse shape," US 9,999,780; DE 10 2010 004 307 B4; EP 2 523 726, Deutsches Patent- und Markenamt (DPMA), U.S. Patent and Trademark Office, European Patent Office, 2010.

[134] S. Goetz, T. Weyh, and H.-G. Herzog, "Circuit topologies and methods for magnetic stimulation," DE 10 2012 101 921 B4, Deutsches Patent- und Markenamt (DPMA), 2012.

[135] S. M. Goetz, "Electronic circuit for magnetic neurostimulation and associated control," DE 10 2017 113 581, Deutsches Patent- und Markenamt (DPMA), 2017.

[136] S. M. Goetz, "Electronic Circuit for Magnetic Neurostimulation and Associated Control," US 2018/0361 168, U.S. Patent and Trademark Office, 2018.

[137] N. Iwamuro and T. Laska, "IGBT History, State-of-the-Art, and Future Prospects," *IEEE Transactions on Electron Devices,* vol. 64, pp. 741-752, 2017.

[138] H. H. Li, M. Trivedi, and K. Shenai, "Dynamics of IGBT performance in hard- and soft-switching converters," in *IAS '95. Conference Record of the 1995 IEEE Industry Applications Conference Thirtieth IAS Annual Meeting*, 1995, pp. 1006-1009, vol. 2. doi: 10.1109/IAS.1995.530411

[139] W. Saito, S. Ono, and H. Yamashita, "Influence of carrier lifetime control process in superjunction MOSFET characteristics," in *2014 IEEE 26th International Symposium on Power Semiconductor Devices & IC's (ISPSD)*, 2014, pp. 87-90. doi: 10.1109/ISPSD.2014.6855982

[140] S. Jahdi, O. Alatise, R. Bonyadi, P. Alexakis, C. A. Fisher, J. A. O. Gonzalez, *et al.*, "An Analysis of the Switching Performance and Robustness of Power MOSFETs Body Diodes: A Technology Evaluation," *IEEE Transactions on Power Electronics,* vol. 30, pp. 2383-2394, 2015.

[141] F. Udrea, G. Deboy, and T. Fujihira, "Superjunction Power Devices, History, Development, and Future Prospects," *IEEE Transactions on Electron Devices,* vol. 64, pp. 713-727, 2017.

[142] J. Fang, F. Blaabjerg, S. Liu, and S. Goetz, "A Review of Multilevel Converters with Parallel Connectivity," *IEEE Transactions on Power Electronics,* vol. 36, pp. 12468-12489, 2021.

[143] D. Freche, J. Naim-Feil, A. Peled, N. Levit-Binnun, and E. Moses, "A quantitative physical model of the TMS-induced discharge artifacts in EEG," *PLOS Computational Biology,* vol. 14, p. e1006177, 2018.

[144] S. M. Goetz, Z. Li, X. Liang, C. Zhang, S. M. Lukic, and A. V. Peterchev, "Control of Modular Multilevel Converter With Parallel Connectivity—Application to Battery Systems," *IEEE Transactions on Power Electronics,* vol. 32, pp. 8381-8392, 2017.

[145] Z. Li, R. Lizana, A. V. Peterchev, and S. M. Goetz, "Distributed balancing control for modular multilevel series/parallel converter with capability of sensorless operation," *IEEE Energ Conv Con and Exp (ECCE),* pp. 1787-1793, 2017. doi: 10.1109/ECCE.2017.8096011

[146] Z. Li, R. Lizana, S. Sha, Z. Yu, A. V. Peterchev, and S. Goetz, "Module Implementation and Modulation Strategy for Sensorless Balancing in Modular Multilevel Converters," *IEEE Transactions on Power Electronics,* vol. 34, pp. 8405-8416, 2018.

[147] Z. Li, R. Lizana, Z. Yu, S. Sha, A. V. Peterchev, and S. Goetz, "A Modular Multilevel Series/Parallel Converter for Wide Frequency Range Operation," *IEEE Transactions on Power Electronics,* vol. 34, pp. 9854-9865, 2019.

[148] Z. Li, J. K. Motwani, Z. Zeng, S. Lukic, A. V. Peterchev, and S. Goetz, "A Reduced Series/Parallel Module for Cascade Multilevel Static Compensators Supporting Sensorless Balancing," *IEEE Transactions on Industrial Electronics*, vol. 68, no. 1, pp. 15-24, 2020.

[149] Z. Li, R. Lizana, A. V. Peterchev, and S. M. Goetz, "Predictive control of modular multilevel series/parallel converter for battery systems," *IEEE Energ. Conv. Con. Exp. (ECCE),* pp. 5685-5691, 2017. doi: 10.1109/ECCE.2017.8096945

[150] S. M. Goetz, T. Weyh, and H. Herzog, "Analysis of a magnetic stimulation system: Magnetic harmonic multi-cycle stimulation (MHMS)," in *2009 International Conference on Biomedical and Pharmaceutical Engineering*, 2009. doi: 10.1109/ICBPE.2009.5384111

[151] S. M. Goetz, D. L. K. Murphy, and A. V. Peterchev, "Magnetic neurostimulation with reduced acoustic emission," US 10,471,272, U.S. Patent and Trademark Office, 2019.